\documentclass{aastex7}

\usepackage{amsmath}

\usepackage[table]{xcolor}

\usepackage{gensymb}
\usepackage{lineno}
\usepackage{soul}
\linenumbers
\usepackage{verbatim}
\usepackage{textcomp}
\usepackage{wasysym}
\def\Rsun{$R_\odot$}
\usepackage{float}
\usepackage[caption = false]{subfig}
\usepackage{paralist}
\usepackage{rotating}
\usepackage{mathrsfs}
\usepackage[color=red!20!white,textsize=tiny]{todonotes}

\usepackage{colortbl}

\newcommand{\unit}[1]{{\text{ #1}}}
\newcommand{\fluxunit}{\unit{counts}\unit{cm}^{-2}\unit{s}^{-1}\unit{sr}^{-1}\unit{(MeV/nuc)}^{-1}}

\begin{document}
\title{Quantifying the Effects of Parameters in Widespread SEP Events with EPREM}
\shorttitle{EPREM Widespread Events}
\correspondingauthor{Matthew A. Young}
\author[0000-0003-2124-7814]{Matthew A. Young}
\email{Matthew.Young@unh.edu}
\affiliation{University of New Hampshire, Durham, NH}
\author[0000-0003-1258-0308]{Bala Poduval}
\email{bala.poduval@unh.edu}
\affiliation{University of New Hampshire, Durham, NH}
\shortauthors{Young and Poduval}

\begin{abstract}
The Energetic Particle Radiation Environment Model (EPREM) solves the focused transport equation (FTE) on a Lagrangian grid in a frame co-moving with the solar wind plasma and simulates the acceleration and transport of solar energetic particles (SEP) in the heliosphere. When not coupled to an external magnetohydrodynamic model, EPREM functions in an uncoupled mode where an ideal cone-shock is injected into a homogeneous background solar wind. We carried out an analysis of the effects of multiple physical parameters in producing widespread SEP events simulated by the uncoupled EPREM using a relatively simple model of a strong magnetized shock propagating radially outward through the inner heliosphere to produce the requisite MHD quantities for EPREM's sophisticated model of proton acceleration and transport. We compared a baseline simulation with seven variations in which the value of a single parameter differed from its baseline value. All simulations exhibit complex profiles of SEP flux as a function of time and energy, with clear dependence on parameters related to diffusion, mean free path, and shock profile. Moreover, while all simulations exhibit significant longitudinal spread in SEP flux, for certain parameter values there exists a decrease or absence in SEP flux at observers located $\geq 90^\circ$ from the shock origin. Relating the differences in SEP flux to the specific values of each parameter in the simulations provides insight into the morphology of observed SEP events and the state of the solar wind through which the driving CME propagates.
\end{abstract}

\section{Introduction}
\label{sec:Introduction}

Energetic particle radiation, including solar energetic particles (SEPs) and galactic cosmic rays (GCRs), is a significant hazard to humans and technological assets throughout the heliosphere. Energetic particles can cause single-event upsets of electronic systems on satellites and can increase the radiation exposure of crewed missions, thereby reducing the ability of crew members to perform certain tasks or undertake additional missions. Exposure to energetic particle radiation can have both short-term and long-term negative health effects \citep{cucinotta_space_2010,cucinotta_space_2014,cucinotta_safe_2015}.

With the advent of the era of multi-spacecraft observations of SEPs have come observations of so-called widespread SEP events. A widespread SEP event is both an intuitive and poorly defined concept. In the most basic sense, widespread SEP events are those in which multiple spacecraft, separated by some distance in longitude, observe SEPs originating from the same source. However, there is currently no agreed-upon longitudinal separation that constitutes a widespread SEP event. This work will defer to the community on the definition of ``widespread'' --- that is, if a peer-reviewed article identifies an SEP event as widespread and provides quantitative justification, we will consider that event to be a widespread SEP event.\added{ See, for example, the description of widespread SEP events within the larger context of multi-spacecraft SEP events in \citet{dresing_solar_2024}.}

Many studies have yielded intriguing questions related to the nature of multi-spacecraft SEP events. For example, widespread multi-spacecraft SEP events have exhibited particle onset times that are not well organized by the spacecraft’s magnetic connection to the CME-driven shock. The most famous of these is the 3~November 2011 event, during which ${>}25$ MeV protons were observed by spacecraft distributed over $360^\circ$ in longitude within 30 minutes of the CME eruption \citep{richardson_identification_2014}.

Smaller $^3$He-rich SEP events have also occasionally been observed over ${>}100^\circ$ despite the presumed solar-flare source, which should be relatively localized \citep{wiedenbeck_observations_2013}. More recently, the ground-level enhancement (GLE) event of 28~October 2021 revealed extraordinarily uniform heavy ion composition over $60^\circ$, something not seen in previous multi-spacecraft composition studies \citep{cohen_longitudinal_2025,guo_first_2023}. In addition to observing SEP events with a wide spread in longitude, the presence of multiple spacecraft aligned along the nominal Parker spiral magnetic field allowed \citet{muro_radial_2025} to quantify radial gradients in H and He fluences, as well as in O and Fe fluences, during the July~17 2023 SEP event.

\citet{dresing_17_2023} analyzed proton and electron fluxes observed at six locations throughout the heliosphere during the 17~April 2021 widespread SEP event. They concluded that a combination of different processes with varying importance at specific locations throughout the inner heliosphere caused the evolution of SEP fluxes observed by spacecraft over a longitudinal range of $210^\circ$. SEPs observed at Parker Solar Probe (PSP) indicated an extended proton injection, likely associated with a shock, as well as an earlier contribution from a flare source. In contrast, the Solar TErrestrial RElations Observatory Ahead (STEREO A) spacecraft and spacecraft located near Earth observed lower-intensity proton fluxes with onset times delayed relative to those at PSP, implying that perpendicular diffusion played a role in transporting protons to those locations, though the authors note that an extended injection region likely supported the widespread SEP event.

This paper presents an analysis of the effects of multiple physical parameters in producing widespread SEP events. It uses a relatively simple model of strong magnetized shock propagating radially outward through the inner heliosphere to produce MHD quantities for a sophisticated model of proton acceleration and transport. Section \ref{sec:EPREM} describes the simulation; section \ref{sec:Simulation Results} presents the results of a baseline simulation run and seven variations, in which the value of one parameter differs from the baseline run; section \ref{sec:Discussion} identifies the salient lessons to be learned by comparing the eight simulation runs; and section \ref{sec:Concluding Remarks} concludes the paper.

\section{EPREM}
\label{sec:EPREM}

The Energetic Particle Radiation Environment Model (EPREM) simulates acceleration and transport of ions throughout the heliosphere by numerically solving the focused transport equation (FTE) on a Lagrangian grid in a frame co-moving with the solar wind plasma. EPREM originated as the Energetic Particle Radiation Environment \emph{Module} within the Earth-Moon-Mars Radiation Environment Model (EMMREM) \citep{schwadron_earth-moon-mars_2010}, where it simulated energetic particle transport for input to radiation dosage models. \citet{kozarev_modeling_2010} used EMMREM to model energetic proton propagation and radiation effectiveness during the 2003 Halloween solar energetic particle (SEP) events. Their work demonstrated the ability of EPREM within EMMREM to simulate energetic particle fluxes that matched those observed by the Ulysses spacecraft, after determining suitable values of parameters that describe the solar-wind plasma.

EPREM subsequently formed a core component of an online tool called Predictions of radiation from REleASE, EMMREM, and Data Incorporating CRaTER, COSTEP, and other SEP measurements (PREDICCS) \citep{schwadron_lunar_2012}. PREDICCS was designed to specify the particle radiation environments of Earth, the Moon, and Mars in near-real-time. Work by \citet{joyce_validation_2013} compared radiation dose rates simulated by PREDICCS to those measured by CRaTER at the Moon during SEP events in January, March, and May of 2012. They found that the dose accumulated in simulations underestimated the CRaTER measurements by as much as 36\% and overestimated it by as much as 10\% --- variations that the authors deemed to be reasonable. A later study by \citet{quinn_modeling_2017} extended the work of \citet{joyce_validation_2013} to comparisons with the Radiation Assessment Detector (RAD) aboard the Mars Science Laboratory (MSL) during MSL's cruise phase to Mars in 2012. They found that the accumulated dose predicted by PREDICCS was also in reasonable agreement with that measured by RAD, with error ranging from 54\% underestimation to 2\% overestimation

Versions of EPREM have been successfully coupled to magnetohydrodynamic (MHD) models, including the Block Adaptive Tree Solar-Wind Roe Upwind Scheme (BATS-R-US) model \citep{kozarev_global_2013}, Enlil \citep{mays_enlil_2016}, and the Magnetohydrodynamic Algorithm outside a Sphere (MAS) \citep{schwadron_particle_2015,linker_coupled_2019}, for the sake of modeling energetic particle transport and acceleration in the simulated MHD environment. 

Recently, \citet{baydin_surrogate_2023} developed a surrogate model of EPREM, called EPREM-S, based on neural networks (NNs) that generates the same output as that of EPREM with great accuracy but tens to hundreds of thousands of times faster. For this, \citet{baydin_surrogate_2023} generated over 30,000 simulations of SEP events using EPREM, where five parameters of the initial seed spectrum were allowed to vary over a range of physically meaningful values, to train and validate EPREM-S and subsequently used for event analysis as a Bayesian inference problem.  

Development of a coupling-agnostic implementation of EPREM began in October of 2020. The term ``coupling-agnostic'' here refers to the planned ability of EPREM to ingest MHD information from an arbitrary source, rather than to be explicitly coupled to a specific MHD model. While that feature is still in development, EPREM is able to internally generate MHD values that represent a simple cone shock, without coupling to any external data source. The EPREM source code, as well as utility scripts and example inputs, are publicly developed in an open-source \texttt{git} repository hosted on GitLab: \url{https://gitlab.com/open-eprem/eprem}. The present study is based on results from v0.14.0 of this ``uncoupled'' implementation of EPREM.

The appendices at the end of this article provide details about components of EPREM relevant to the work presented here: Appendix \ref{sec:Focused Transport Equation} describes how EPREM solves the focused transport equation, Appendix \ref{sec:Lagrangian Grid} describes the EPREM grid of nodes, Appendix \ref{sec:Ideal Shock} describes how EPREM internally generates an ideal shock for accelerating ions, Appendix \ref{sec:Observers} describes the types of observers in simulation runs, and Appendix \ref{sec:Background Spectrum} describes the background flux spectrum that EPREM simulations use.

\section{Simulation Results}
\label{sec:Simulation Results}

This work presents 8 EPREM simulation runs with the ideal shock model (cf. \S~\ref{sec:Ideal Shock}). Table \ref{tab:common-parameters} lists the values of physical parameters common to all EPREM simulation runs presented in this work. All simulation runs used 294 simulation streams, which provided a reasonable balance between simulation run time and resolution when interpolating physical quantities from Lagrangian nodes to Eulerian point-observer locations. Each stream comprised 2000 nodes, which sets the initial outer ``boundary'' at approximately 3.5~au (cf. \S~\ref{sec:Lagrangian Grid}). 

\begin{figure}
    \centering
    \includegraphics[width=1.0\linewidth]{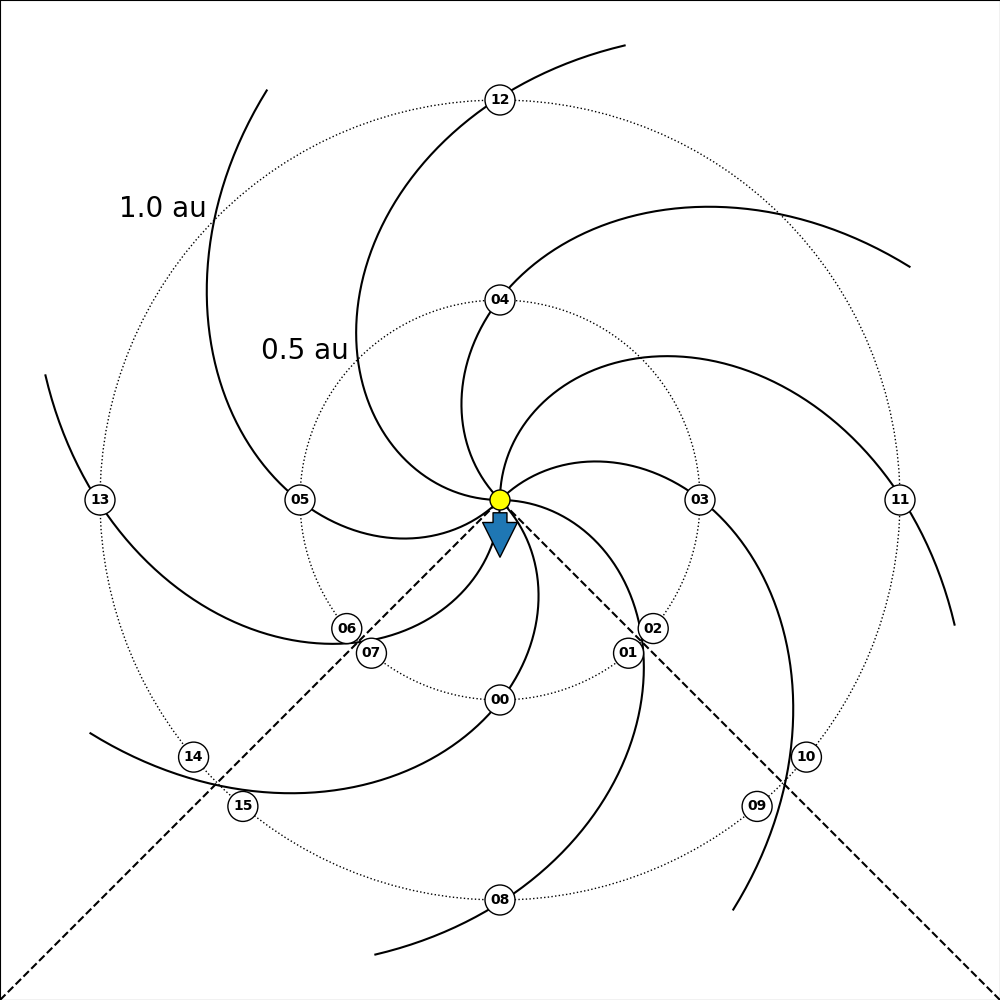}
    \caption{Positions of point observers. Dotted circles indicate observer radii. The small circle at the center represents the Sun and the arrow protruding from the Sun indicates the origin and direction of the ideal shock.\added{ Solid black lines trace out the nominal Parker spiral.} Dashed lines show the region of influence of the ideal shock.}
    \label{fig:observer-positions}
\end{figure}

All simulation runs also included 16 point observers: 8 at 0.5~au and 8 at 1.0~au. Observer azimuthal locations were chosen to cover a wide range inside and outside of the shock cone. At each radial distance, there are point observers at $0^\circ$, $+90^\circ$, $180^\circ$, and $-90^\circ$ from the shock origin, as well as point observers $\pm 5^\circ$ from each edge of the shock cone.\added{ The sign convention for angles is such that positive angles are west of the shock origin and negative angles are east of the shock origin.} All point-observer locations are in the solar equatorial plane. See \S~\ref{sec:Observers} for descriptions of EPREM's stream and point observers. Figure \ref{fig:observer-positions} displays the radial and azimuthal locations of these point observers. Each observer's numerical label is arbitrary but will be useful for later reference. The small yellow circle at the center represents the location of the Sun, as well as the approximate location of the inner simulation boundary. The blue arrow protruding from the Sun represents the origin of the ideal shock and the direction in which it propagates. Black dashed lines extending from the nominal solar surface indicate the region over which the outwardly propagating ideal shock extends --- point observers within this region record the shocked MHD quantities while those outside do not. Finally, black dotted lines indicate the radii of point observers. Descriptions of individual simulated SEP events will at times refer to ``observable'' flux values. By ``observable'' in this context, we mean an amplitude larger than $10^{-5}$ counts $\left(\text{cm}^{2}\unit{s}\unit{sr}\unit{MeV}\right)^{-1}$, which we have chosen in order to be somewhat consistent with typical instrument ranges.\added{ Note that, though this setup allows for some crude analysis of radial variation in SEP fluxes, this work will focus only on the longitudinal variation. Future work may include similar analysis of radial variation.}

See Appendix \ref{sec:Ideal Shock} for a description of the analytic form of the ideal shock. When computing the shocked density, the value of \texttt{idealShockJump} sets the compression ratio, $q_{s}$, and the value of \texttt{mhdDensityAu} sets the reference density, $n_{0}$. When computing the shocked radial velocity component, the value of \texttt{flowMag} sets the unshocked solar-wind speed, $V_{r1}$, and the value of \texttt{idealShockSpeed} sets the shock speed, $V_{s}$. When computing the shocked azimuthal magnetic-field component, the value of \texttt{mhdBAu} sets the reference magnetic field, $B_{0}$. At the shock front, \texttt{idealShockGradient} controls the deviation from a step function by setting the value of $\Gamma_{s}$. Downstream of the shock, \texttt{idealShockFalloff} sets the value of $\xi_{s}$. All simulation runs described here use a value of $\xi_{s} = 1.0$, whereas setting $\xi_{s} = 0.0$ (the current default behavior) would produce asymptotically constant downstream MHD quantities.

Appendix \ref{sec:Background Spectrum} describes the flux spectrum used to initialize the ion distribution(s), $f_{s}$, on each simulation node. The initial spectrum thus defines the upstream, or background, population through which the ideal shock propagates. All simulation runs presented here used $J_{0} = 10\ \left(\text{cm}^{2}\unit{s}\unit{sr}\unit{MeV}\right)^{-1}$, $r_{0} = 1\unit{au}$, \added{$\beta = 2$}, $E_{0} = 50\unit{keV}$, $\gamma = 2$, and $E_{c} = 1\unit{MeV}$, and exclusively modeled energetic protons.

Given the values of \texttt{idealShockInitTime} ($t_{s}$) and \texttt{idealShockSpeed} ($V_{s}$) listed in Table \ref{tab:common-parameters}, the ideal shock reaches the radial distance of the initial outermost node, $R_{\text{max}}$, in 6 days. We therefore chose to end each simulation run after 6 days of simulated time.

In addition to the parameters listed in Table \ref{tab:common-parameters}, all simulation runs used seven variable parameters. Each simulation run except for a baseline run varied the value of one such parameter while holding all others fixed.

\begin{table}[!h]
    \centering
    \begin{tabular}{|r|c|l|l|c}
        \hline
        Name & Symbol & Description & Value \\
        \hline
        \texttt{tDel} & $\delta t$ & stimulation time-step length & 0.01 day \\
        \texttt{eMin} & $E_{\text{min}}$ & lowest simulated ion energy & 0.050 MeV \\
        \texttt{eMax} & $E_{\text{max}}$ & highest simulated ion energy & 500.0 MeV \\
        \texttt{mhdDensityAu} & $n_{0}$ & solar-wind density at 1 au & 5 $\text{cm}^{-3}$ \\
        \texttt{mhdBAu} & $B_{0}$ & magnitude of solar-wind magnetic field at 1 au & 5 nT \\
        \texttt{flowMag} & $V_{r}$ & radial speed of background (unshocked) solar wind & 300.0 km/s \\
        \texttt{boundaryFunctAmplitude} & $J_{0}$ & background flux of 50-keV protons & 10 $\left(\text{cm}^{2}\unit{s}\unit{sr}\unit{MeV}\right)^{-1}$ \\
        \texttt{boundaryFunctR0} & $r_{0}$ & reference radial distance for background spectrum & 1.0 au \\
        \texttt{boundaryFunctBeta} & $\beta$ & radial power-law index of background spectrum & 2 \\
        \texttt{boundaryFunctE0} & $E_{0}$ & reference energy for background spectrum & 50 keV \\
        \texttt{boundaryFunctGamma} & $\gamma$ & energetic power-law index of background spectrum & 4 \\
        \texttt{boundaryFunctEcutoff} & $E_{c}$ & roll-over energy of background spectrum & 1.0 MeV \\
        \texttt{idealShockInitTime} & $t_{s}$ & launch time of ideal shock & 1.0 day \\
        \texttt{idealShockTheta} & $\Theta_{0}$ & polar coordinate of ideal-shock origin & $90.0^\circ$ \\
        \texttt{idealShockPhi} & $\Phi_{0}$ & azimuthal coordinate of ideal-shock origin & $0.0^\circ$ \\
        \texttt{idealShockThetaWidth} & $\Delta\Theta$ & polar opening angle of ideal shock & $90.0^\circ$ \\
        \texttt{idealShockPhiWidth} & $\Delta\Phi$ & azimuthal opening angle of ideal shock & $90.0^\circ$ \\
        \texttt{idealShockJump} & $q_{s}$ & relative density increase across ideal shock & 4.0 \\
        \texttt{idealShockSpeed} & $V_{s}$ & outward radial speed of ideal shock & 1200 km/s \\
        \texttt{idealShockFalloff} & $\xi_{s}$ & exponential relaxation rate of shocked quantities & 1.0 \\
        \hline
    \end{tabular}
    \caption{Parameter values common to all simulation runs.}
    \label{tab:common-parameters}
\end{table}

\begin{table}[!h]
    \centering
    \begin{tabular}{|r|c|l|}
        \hline
        Name & Symbol & Description \\
        \hline
        \texttt{lamo} & $\lambda_{\parallel 0}$ & mean free path of a proton with 1 GV rigidity at 1 au \\
        \texttt{mfpPower} & $\beta$ & power-law index of mean free path \\
        \texttt{mfpInverseB} & (none) & if true, $\lambda_{\parallel} \propto |\vec{B}|^{-\beta/2}$; otherwise, $\lambda_{\parallel} \propto r^{\beta}$ \\
        \texttt{rigidityPower} & $\chi$ & mean-free-path power-law dependence on rigidity \\
        \texttt{idealShockGradient} & $\Gamma_{s}$ & similarity of the shock front to a step function \\
        \texttt{kperxkpar} & $\kappa_{\perp}/\kappa_{\parallel}$ & ratio of perpendicular to parallel diffusion coefficients \\
        \hline
    \end{tabular}
    \caption{Parameters that vary across simulation runs.}
    \label{tab:parameters}
\end{table}
\clearpage

\begin{table}[!h]
    \centering
    \begin{tabular}{|c c c c c c c|}
        \hline
        Run Name & $\kappa_{\perp}/\kappa_{\parallel}$ & $\beta$ & $\lambda_{\parallel 0}$ & $\chi$ & $\lambda_{\parallel} \propto \{|\vec{B}|^{-\beta/2}, r^{\beta}\}$ & $\Gamma_{s}$ \\
        \hline\hline
        baseline & $10^{-2}$ & 2 & 0.1~au & 1/3 & $\lambda_{\parallel} \propto |\vec{B}|^{-\beta/2}$ & 100 \\
        \hline
        \added{no-perp-diff} & \cellcolor{lightgray}0.0 & 2 & 0.1~au & 1/3 & $\lambda_{\parallel} \propto |\vec{B}|^{-\beta/2}$ & 100 \\
        \hline
        mfp-power-hi & $10^{-2}$ & \cellcolor{lightgray}4 & 0.1~au & 1/3 & $\lambda_{\parallel} \propto |\vec{B}|^{-\beta/2}$ & 100 \\
        \hline
        mfp-power-lo & $10^{-2}$ & \cellcolor{lightgray}1 & 0.1~au & 1/3 & $\lambda_{\parallel} \propto |\vec{B}|^{-\beta/2}$ & 100 \\
        \hline
        mfp-reference & $10^{-2}$ & 2 & \cellcolor{lightgray}1.0~au & 1/3 & $\lambda_{\parallel} \propto |\vec{B}|^{-\beta/2}$ & 100 \\
        \hline
        mfp-rigidity & $10^{-2}$ & 2 & 1.0~au & \cellcolor{lightgray}2/3 & $\lambda_{\parallel} \propto |\vec{B}|^{-\beta/2}$ & 100 \\
        \hline
        mfp-scaling & $10^{-2}$ & 2 & 0.1~au & 1/3 & \cellcolor{lightgray}$\lambda_{\parallel} \propto r^{\beta}$ & 100 \\
        \hline
        shock-profile & $10^{-2}$ & 2 & 0.1~au & 1/3 & $\lambda_{\parallel} \propto |\vec{B}|^{-\beta/2}$ & \cellcolor{lightgray}10 \\
        \hline
    \end{tabular}
    \caption{Variations in parameter values across all simulation runs. The parameter values modified from the baseline run are indicated by gray cells.}
    \label{tab:variable-parameters}
\end{table}

\subsection{Baseline Simulation Run}
\label{sec:Baseline Simulation Run}

The ``baseline'' simulation run was designed to be a relatively realistic starting point for probing the physical processes that produce widespread SEP events. In other words, this baseline simulation run does not use strictly default parameter values nor is it a simulation run in which nothing interesting happens.

\begin{figure}
    \centering
    \includegraphics[width=0.75\linewidth]{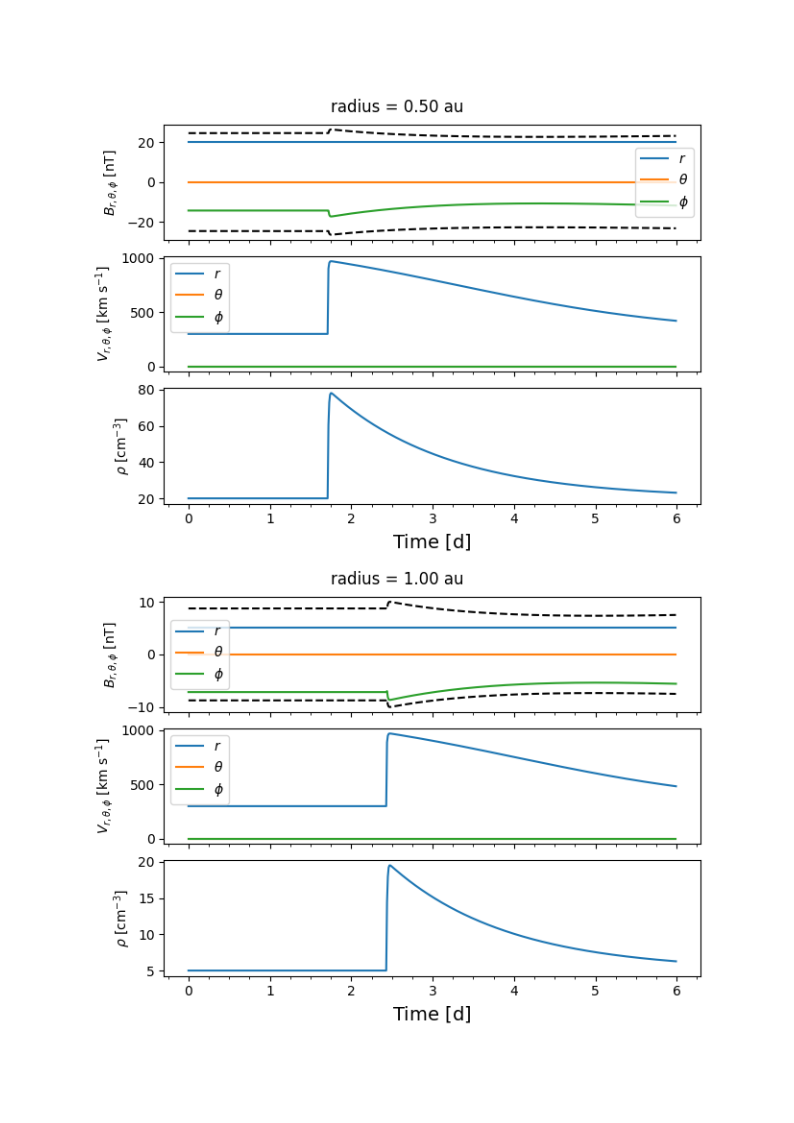}
    \vspace{-5em}
    \caption{MHD quantities as functions of time during the baseline simulation run. The top set of panels shows components of the magnetic field, the velocity field, and the density at 0.5~au. The bottom set of panels show the same quantities at 1.0~au. Black dashed lines bounding the magnetic-field components represent $\pm |\vec{B}|$.}
    \label{fig:observer-mhd-baseline}
\end{figure}

The baseline values of \texttt{mhdDensityAu}, \texttt{mhdBAu}, \texttt{flowMag}, \texttt{idealShockJump}, \texttt{idealShockSpeed}, and \texttt{idealShockGradient} produce the plots of MHD quantities shown in Figure \ref{fig:observer-mhd-baseline}. The top set of panels show the radial, polar, and azimuthal coordinates of the magnetic field, $\vec{B}$, the equivalent components of the velocity field, $\vec{V}$, and the density, $n$, for an observer inside the shock cone at 0.5~au. The bottom set of panels show the same quantities for an observer inside the shock cone at 1.0~au. The simplistic implementation of the EPREM ideal shock causes all observers inside the shock cone to record the same values of MHD quantities over time. Similarly, all observers outside the shock cone record only the unshocked values of MHD quantities (not shown).

The radially propagating shock front should arrive at radial distance $R_{*}$ at time $t_{*} = t_{s} + \left(R_{*}/V_{s}\right)$, where $t_{s}$ and $V_{s}$ have the meanings given in Table \ref{tab:common-parameters}. Plugging in appropriate values gives arrival times of approximately 1.72 days at 0.5~au and 2.45 days at 1.0~au, which are consistent with the time profiles shown in Figure \ref{fig:observer-mhd-baseline}.\added{ Due to the use of a Lagrangian grid, in which computational nodes follow the local fluid flow, the shape of streams near the ideal shock trace out its propagation. The evolution of the ideal-shock cone is therefore visible in figures such as Figure \ref{fig:stream-flux-baseline}.}

\begin{figure}
    \centering
    \includegraphics[width=1.0\linewidth,trim={2cm 2cm 2cm 2cm},clip=true]{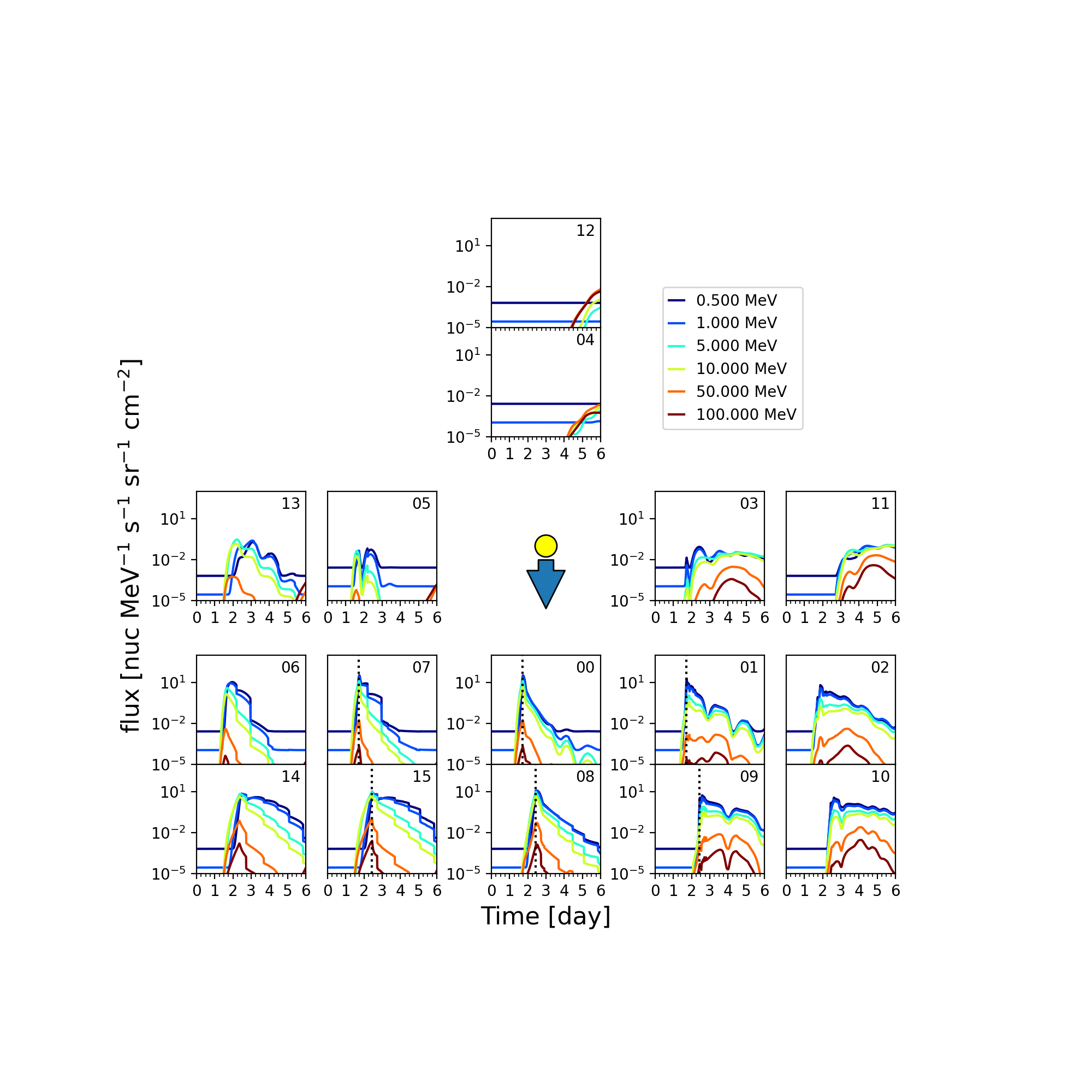}
    \vspace{-5em}
    \caption{Flux at each point observer during the baseline simulation run. Observer locations have been adjusted to avoid overlap and to condense the figure. Refer to Figure \ref{fig:observer-positions} for correspondence between observer number and location. Vertical dotted lines indicate the time of shock passage, where applicable. All panels share a common x axis, labeled below panel 08, and y axis, labeled to the left of panel 13.}
    \label{fig:observer-flux-baseline}
\end{figure}

\begin{figure}
    \centering
    \includegraphics[width=1.0\textwidth]{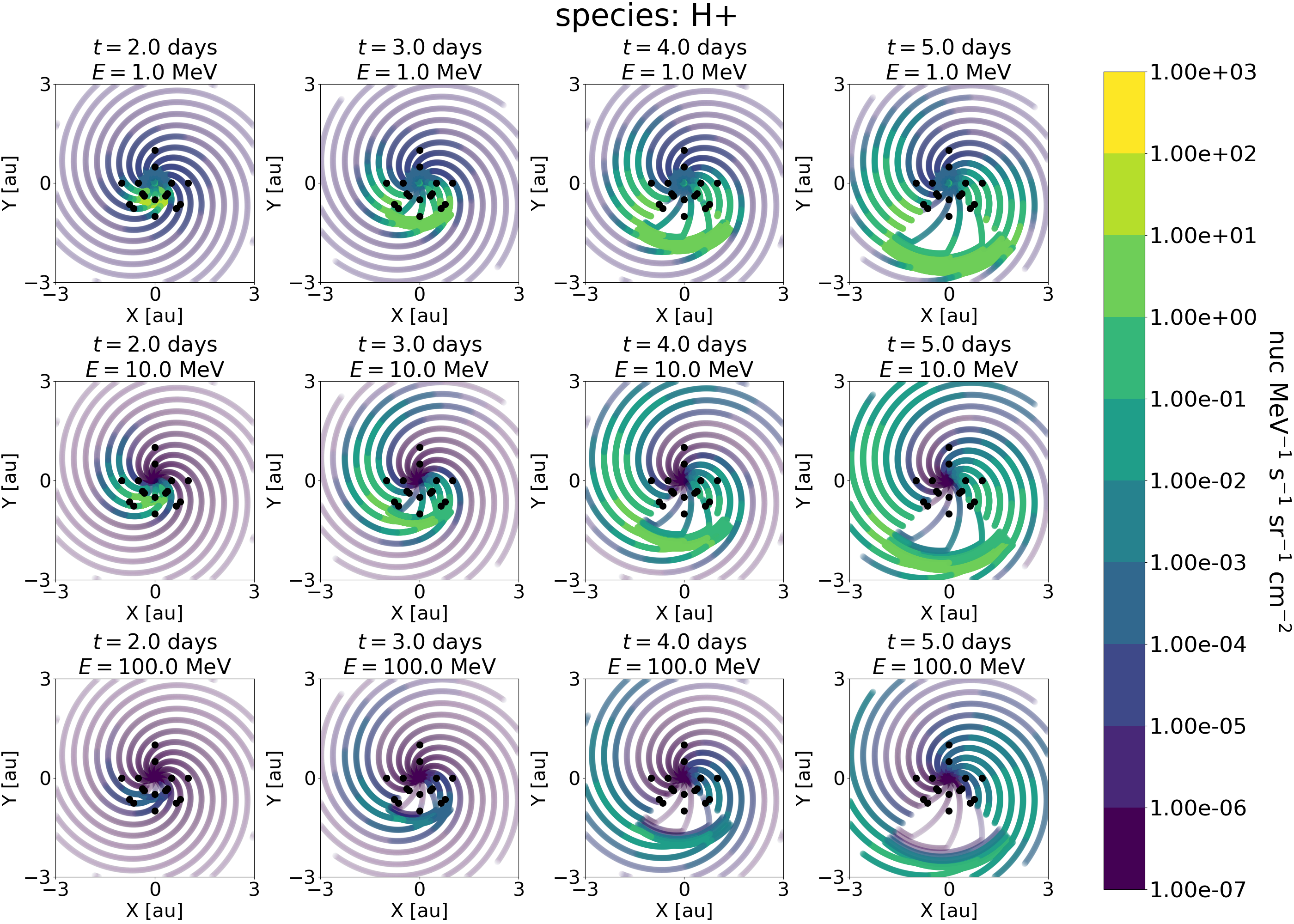}
    \caption{Fluxes of protons with energies (top-to-bottom) 1~MeV, 10~MeV, and 100~MeV at simulation days (left-to-right) 2, 3, 4, and 5\added{ during the baseline simulation run, on a subset of streams in the solar ecliptic plane}. Point observer locations are indicated by black circles. Note that the location of color-coded stream nodes trace out the ideal-shock cone due to the nature of the Lagrangian grid.}
    \label{fig:stream-flux-baseline}
\end{figure}

Figure~\ref{fig:observer-flux-baseline} shows simulated flux of protons with energies of 0.5, 1, 5, 10, 50, and 100 MeV at each observer during the baseline simulation run. The vertical dotted line in panels labeled 00, 01, 07, 08, 09, and 15 indicate the time at which the ideal shock passed that observer. The first point to note, from a big-picture perspective, is that substantial flux at all energies reaches most observers outside the ideal-shock cone (cf. Figure~\ref{fig:observer-positions}). Only the pair of observers separated by $180^\circ$ from the ideal-shock origin (panels 04 and 12) fail to record any increased flux associated with shock passage. However, both of those observers still record increased flux starting at day 4 or 5. This increased flux appears to be the result of\added{ proton diffusion perpendicular to the magnetic field, as will be discussed in subsequent sections}. An extended baseline simulation run (not included here), which ran to 12 days, suggests that fluxes decrease after these late post-shock increases.

Figure~\ref{fig:stream-flux-baseline} shows simulated flux of protons with 1, 10, and 100 MeV on a subset of streams at 2, 3, 4, and 5 days into the baseline simulation run. The panels of 10-MeV and 100-MeV proton flux at day 5 \added{demonstrate the effect of perpendicular diffusion in radially spreading accelerated protons across magnetic field lines, leading to the flux increases at $180^\circ$ from the shock origin near the end of the simulation run}.

The observers $5^\circ$ inside the western edge of the shock \added{cone} record the most abrupt flux increases. After the shock passes, simulated proton fluxes broadly decay until the end of the simulation. The observer at 0.5~au records an overall decay of protons with 1-10 MeV, while protons with 50 and 100 MeV experience more of a plateau until day 4, at which point they quickly decay below observable values. The observer at 1.0~au records a flux depletion at all energies around day 4, after which all fluxes decay toward background (or at least unobservable) values. Fluxes of protons with 1-10 MeV steadily decay before and after the depletion; fluxes of protons with 50 and 100 MeV rises toward the depletion, then quickly decay afterward.

The observers $5^\circ$ outside the western edge also record abrupt flux increases near the time of shock passage, even though the shock itself does not pass over their locations. One possible explanation could be that those observers are simply interpolating flux from nodes that \emph{are} inside the shock cone, but images of the flux on nearby streams strongly suggest that this is not the case (cf. Figure~\ref{fig:stream-flux-baseline}).\added{ One component of the explanation for these flux spikes is perpendicular diffusion of accelerated particles beyond the shock cone. Refer to \citet{axford_modulation_1965,parker_passage_1965} for derivations of the classical scattering expression for $\kappa_{\perp}/\kappa_{\parallel}$, in addition to more recent theoretical and numerical studies by \citet{jokipii_rate_1987,matthaeus_spatial_1995,giacalone_transport_1999,matthaeus_nonlinear_2003,zank_particle_2006}. \added{Such diffusion} would account for the small delay between the peak at each observer $5^\circ$ inside the shock cone and the corresponding observer $5^\circ$ outside the shock cone. Another component of the explanation is the similar magnetic connectivity between observers 01 and 02, and 09 and 10 (cf. Figure \ref{fig:observer-positions}).}

Pairs of observers on either side of the eastern edge of the shock cone record flux profiles that are more similar than pairs of observers on either side of the western edge. As with the western-edge observers, there is a reasonable physical explanation, beyond overly aggressive interpolation, for the presence of shock-related flux increases at the locations of observers outside the shock cone. In this case, the observers are well connected to the location on the inner boundary where the shock originates, allowing protons accelerated at lower radial distances to propagate along streams to their locations. Examining the peak amplitudes of 10-MeV protons in Figure \ref{fig:observer-flux-baseline} and the location of increased 10-MeV flux in Figure \ref{fig:stream-flux-baseline} suggests that more protons propagate \added{parallel to magnetic field lines} to reach the locations of observers $5^\circ$ outside the \emph{eastern} edge of the shock cone than \added{diffuse across field lines} to reach the locations of observers $5^\circ$ outside the \emph{western} edge of the shock cone.

Moving farther away from the shock cone, we see greater differences both between and among pairs of observers. The pair of observers at $-90^\circ$ from (i.e., $90^\circ$ east of) the shock origin exhibit increased fluxes associated with the shock passage on connected streams, but the arrival times are delayed due to propagation from the acceleration region. The 1.0-au observer is well enough connected to the shock front and the shocked region that it records extended elevated fluxes of protons with energies up to \added{50} MeV. The 0.5-au observer records an initial peak in protons with energies up to 50 MeV, but the high-energy protons quickly stream past and, though the shocked region is able to produce a secondary peak in protons with energies up to 10 MeV, this observer does not record significant fluxes of \added{100-MeV} protons until \added{partway through day 5} of the simulation.\added{ This late increase is most likely due to perpendicular diffusion.} One final point of interest is that the peak in 1-MeV protons at 1.0~au lags the peak in higher-energy protons by at least a day. This appears to be due to a combination of delayed acceleration at nodes on the eastern edge of the shock cone, which later connect to that observer's location, and the time necessary to propagate to that location after the streams have rotated.

Fluxes recorded by the observers at $+90^\circ$ from (i.e., $90^\circ$ west of) the shock origin exhibit behaviors more strongly related to \added{perpendicular diffusion}. The initial spike of protons with 5 and 10 MeV recorded by the 0.5-au observer is likely due to a portion of nodes on the western edge of the shock cone briefly connecting to that location as the shock is beginning to accelerate protons to those energies. As the shock moves outward, it produces much larger fluxes of 5-MeV and 10-MeV protons, some of which \added{diffuse radially outward} past both observer locations --- first at 0.5~au, then at 1.0~au. The 0.5-au observer records a small increase in 50-MeV protons, followed by a larger increase in both 50-MeV and 100-MeV protons accelerated later in the shock's radial evolution. The 1.0-au observer records larger corresponding increases that arrive almost a day later\added{. Comparisons to simulation runs without perpendicular diffusion and with neither perpendicular nor parallel diffusion, described in \S \ref{sec:The Role of Perpendicular Diffusion}, suggests that perpendicular diffusion produces the initial high-energy increase (e.g., in 50-MeV protons at observer~03 during day~2) while parallel diffusion produces the later increases (e.g., in 50-MeV and 100-MeV protons at observer~03 from day~3 onward).}

\subsection{The Role of Perpendicular Diffusion}
\label{sec:The Role of Perpendicular Diffusion}

\begin{figure}
    \centering
    \includegraphics[width=1.0\linewidth,trim={2cm 2cm 2cm 2cm},clip=true]{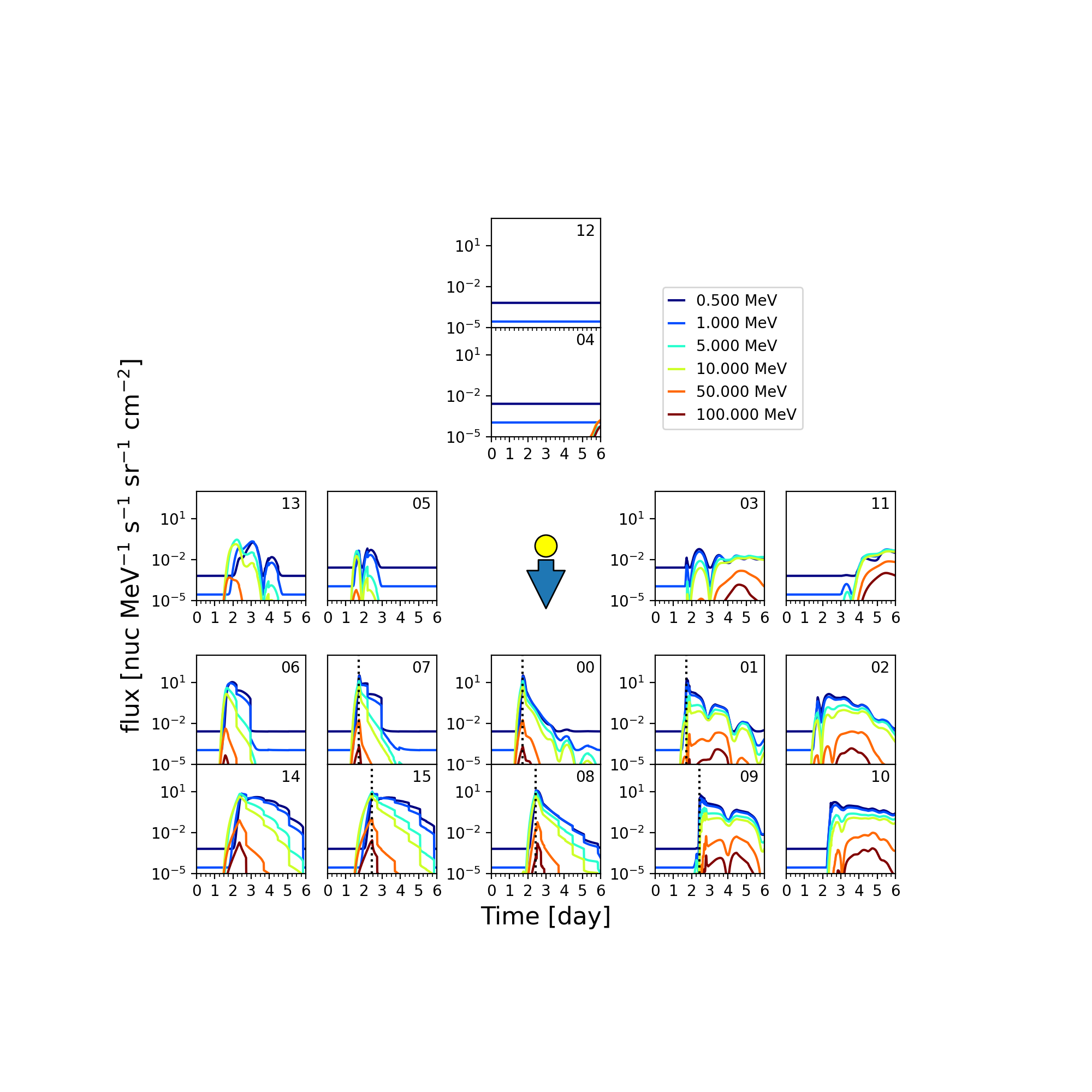}
    \vspace{-5em}
    \caption{Flux at each point observer during the simulation run with no perpendicular diffusion. The layout is identical to that of Figure~\ref{fig:observer-flux-baseline}.}
    \label{fig:observer-flux-no-perp-diff}
\end{figure}

\begin{figure}
    \centering
    \includegraphics[width=1.0\textwidth]{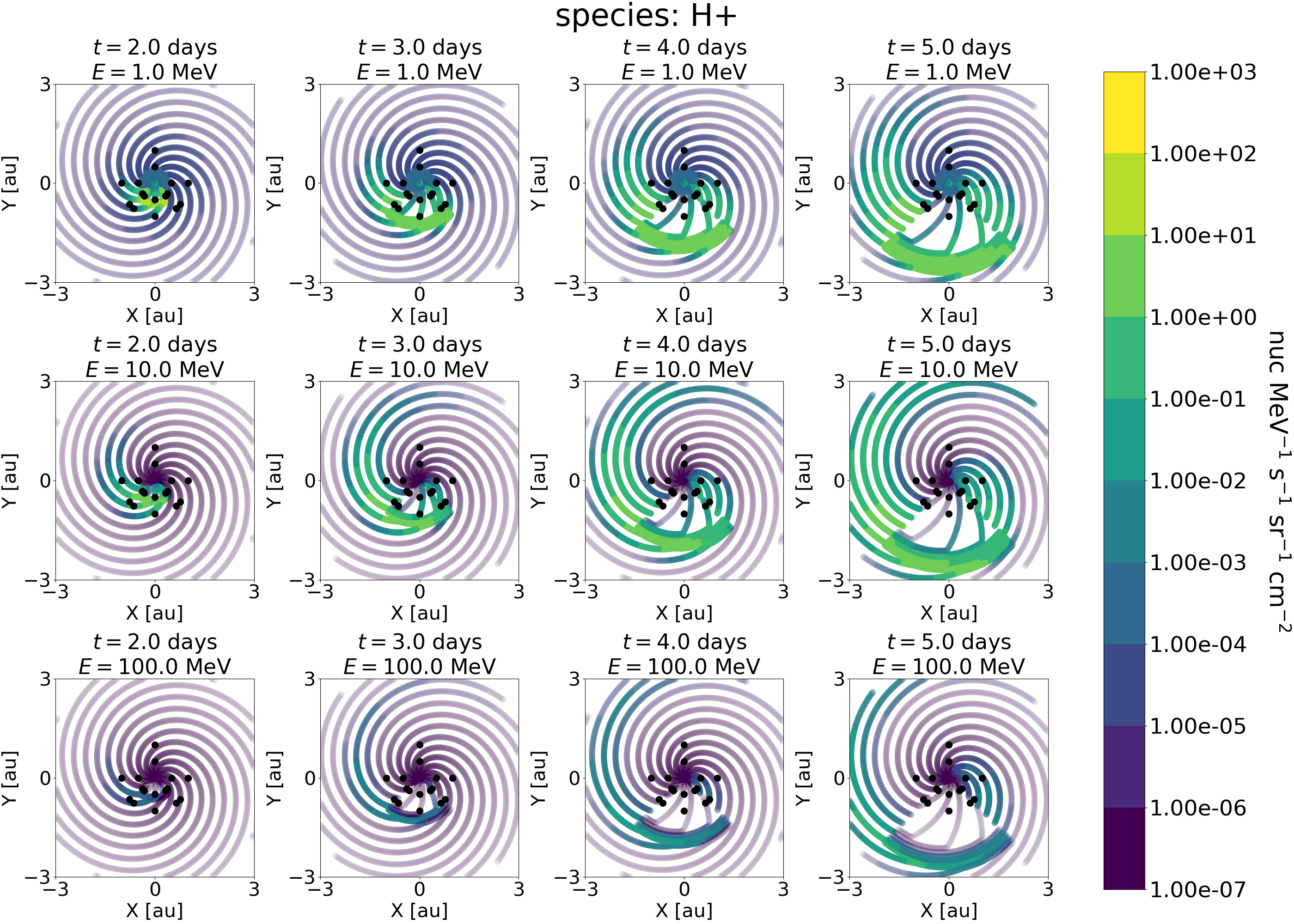}
    \caption{Fluxes of protons during the simulation run with no perpendicular diffusion. The layout is identical to that of Figure~\ref{fig:stream-flux-baseline}.}
    \label{fig:stream-flux-no-perp-diff}
\end{figure}

To examine the role that perpendicular diffusion plays in transporting flux throughout the heliosphere, we ran a variation of the baseline simulation without perpendicular diffusion by setting $\kappa_{\perp}/\kappa_{\parallel} = 0$. Figures~\ref{fig:observer-flux-no-perp-diff} and \ref{fig:stream-flux-no-perp-diff} show proton fluxes from this simulation run. A noticeable (though not surprising) result of suppressing perpendicular diffusion is the near absence of flux increases at the 0.5-au point observer located at $180^\circ$ from the shock origin, and the complete absence of features at the corresponding 1.0-au point observer.\added{ This result supports the suggestion in \S \ref{sec:Baseline Simulation Run} that perpendicular diffusion is responsible for the flux increases recorded by observers 04 and 12 in the baseline simulation run.}

On the opposite side of the Sun, both point observers at the center of the shock cone recorded fluxes that differ very little from their values in the baseline simulation run (cf. \ref{fig:observer-flux-baseline}), except for a slightly more abrupt pre-shock rise at 50 and 100~MeV, and 50~MeV flux that falls below $10^{-5}$ before day~4.

These behaviors broadly apply to the point observers that straddle the eastern edge of the shock cone as well: Although the pre-shock flux profiles differ little from their baseline values, fluxes of protons with 50 and 100 MeV decay slightly faster at the 1.0-au point observers while fluxes of protons with energies $\ge 1$ MeV decay more quickly at the 0.5-au observer than in the baseline simulation run. Finally, the increased flux of 50-MeV and 100-MeV protons at the very end of the simulation run seems to have disappeared with the lack of perpendicular diffusion.

This trend continues to the 0.5-au observer at $-90^\circ$, but the deviation from baseline flux profiles is significantly more drastic at the corresponding 1.0~au observer. Not only does that observer record faster post-shock decay in fluxes of protons with 50 and 100 MeV, but the flux of 5~MeV and 10~MeV protons completely drops out at approximately 3.5~days, then briefly rebounds to a fraction of its peak before falling back below observable values approximately 24~hours later. The flux of protons with $\leq 1\unit{MeV}$ mirrors this behavior, but the resulting difference in amplitude around day~4 is not as drastic. Besides the loss of a small bump in the flux of 1-MeV protons recorded by the 0.5~au observer on day 3, the other notable effect at $-90^\circ$ of suppressing perpendicular diffusion is again the absence of high-energy protons during day 5.

Fluxes of protons with $\le 10\unit{MeV}$ recorded by the point observers inside the western edge of the shock cone differ only slightly from the corresponding baseline profiles, with the exception of a sharper shock-associated peak observed at 0.5~au. The differences in fluxes of 50-MeV and 100-MeV protons are more pronounced: Both profiles exhibit much sharper initial peaks than in the baseline simulation run, and, with the exception of 50-MeV protons at 1.0~au, the depletions around day 4 are deeper. Just outside the western edge of the shock cone, the lack of perpendicular diffusion noticeably alters fluxes at both observers, with the most significant differences recorded by the observer at 0.5~au. That observer records a relatively sharp peak in fluxes of protons with $\le 10\unit{MeV}$ around the time when the shock passed $5^\circ$ east of its location, then a brief depletion that was followed by a rise to higher fluxes than recorded during the initial peak. The flux of 50-MeV protons exhibited pseudo-oscillatory behavior before rising to values consistent with the corresponding baseline post-shock profile, and the flux of 100-MeV protons appeared to exhibit similar behavior, though the initial peak either does not exist or is below observable values. The observer at 1.0~au and $5^\circ$ west of the shock cone recorded fluxes of protons with $\le 10\unit{MeV}$ that did not meaningfully differ from baseline values. However, the fluxes of 50-MeV and 100-MeV protons not only exhibit more pronounced depletions around day 3, but also rise to their initial peak approximately 6~hours later than in the baseline simulation run.

At $+90^\circ$ from the shock origin, the 0.5-au observer records a strong depletion around day 3 in place of the short dip recorded in the baseline simulation run, followed by relatively lower fluxes of 50-MeV and 100-MeV protons. The differences from baseline fluxes at the 1.0-au observer are even more pronounced, in the sense that fluxes of 50-MeV and 100-MeV protons do not increase to observable levels until day 4, while fluxes of 5-MeV and 10-MeV protons experience a short bump before rising to their baseline levels.

Figure \ref{fig:stream-flux-no-perp-diff} illustrates how suppressing perpendicular diffusion removes a substantial amount of flux at all energies from streams that connect to the observers $180^\circ$ from the shock origin. It also suggests that perpendicular diffusion of higher-energy protons\added{ causes the increases in 50-MeV and 100-MeV proton fluxes recorded by the observers $+90^\circ$ from the shock origin during days 2--3 of the baseline simulation run, which are evident in Figure \ref{fig:observer-flux-baseline} but not in Figure \ref{fig:observer-flux-no-perp-diff}}. Finally, it shows how perpendicular diffusion sustains fluxes, especially of lower-energy protons, on streams connected to eastern observers.\added{ A simulation run with neither perpendicular nor parallel diffusion (not shown) produced no observable fluxes of protons with energies above 10~MeV, and observers $+90^\circ$ west of the shock origin recorded no flux increases (at any energy) after day 3. In fact, observer 11 (at $+90^\circ$ and 1.0 au) did not record any evidence of the shock passage in that simulation run. This suggests that flux increases at $+90^\circ$ later in the simulation run without perpendicular diffusion result from parallel diffusion along field lines that were initially connected to the shock cone, then became connected to observers 03 and 11. It is likely that these fluxes represent back-streaming protons --- that is, protons that propagated sunward along field lines connected to observers 03 and 11 after having been accelerated by the shock. Although it is not possible to determine proton anisotropies from the pre-computed flux output used in this work, future simulation runs that output the full particle distribution will allow us to test this claim and further examine the role of back-streaming protons in contributing to widespread SEP events.}

\subsection{The Role of Mean Free Path}
\label{sec:The Role of Mean Free Path}

EPREM computes the parallel scattering mean free path, $\lambda_{\parallel}$, at each node during each time step via one of the following equations
\begin{subequations}
\begin{eqnarray}
    \lambda_{\parallel}(r) &=& \lambda_{\parallel 0}\left(\frac{\mathcal{R}}{\mathcal{R}_{0}}\right)^{\chi} \left(\frac{r}{\text{1 au}}\right)^{\beta} \qquad \left(\texttt{mfpInverseB=0}\right) \\
    \lambda_{\parallel}(|\vec{B}|) &=& \lambda_{\parallel 0}\left(\frac{\mathcal{R}}{\mathcal{R}_{0}}\right)^{\chi} \left(\frac{|\vec{B}|}{B_{0}}\right)^{-\beta/2} \quad \left(\texttt{mfpInverseB=1}\right)
\end{eqnarray}
\end{subequations}

\noindent where $\mathcal{R} = pc/q$ is the rigidity of an ion with charge $q$ and momentum $p$, $\mathcal{R}_{0} \equiv m_{0}v_{0}c / q_{0}$ is a reference rigidity at speed $v_{0}$ for an ion with mass $m_{0}$ and charge $q_{0}$, and $B_{0}$ is the magnitude of the magnetic field at 1.0~au as set by the parameter \texttt{mhdBAu}. The default value in EPREM of the rigidity power-law index, $\chi$, is $1/3$; actual values based on fits to observations may vary in the range $0.3 \le \chi < 1$ \citep{droge_rigidity_2000,droge_probing_2005}.

To examine the role that the mean free path plays in SEP acceleration and transport, we ran five variations of the baseline simulation: one with $\lambda_{\parallel 0} = 1.0\unit{au}$, one with $\chi = 2/3$, one with $\beta = 4.0$, one with $\beta = 1.0$, and one with $\lambda(r) \propto r^{2}$.

\begin{figure}
    \centering
    \includegraphics[width=1.0\linewidth,trim={2cm 2cm 2cm 2cm},clip=true]{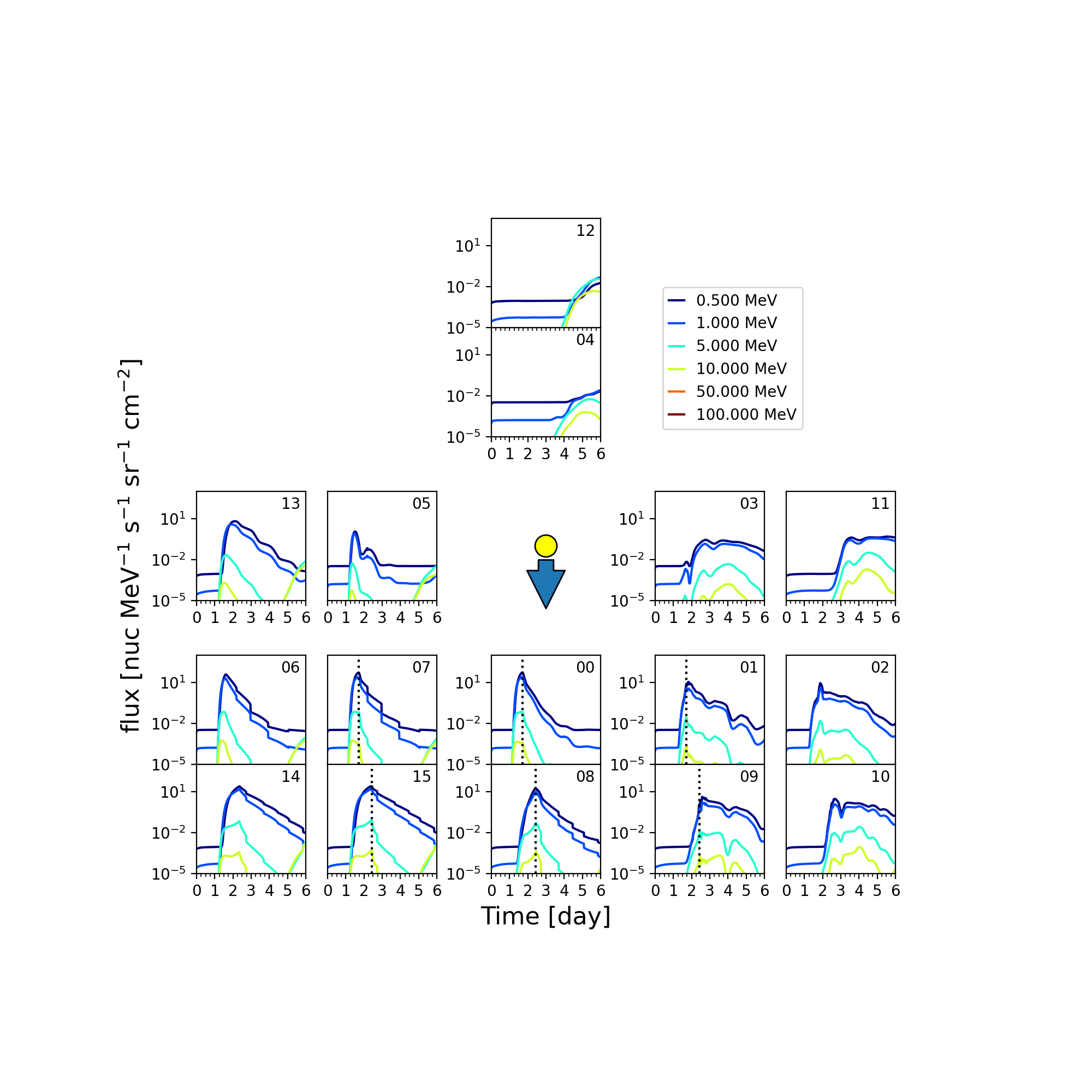}
    \vspace{-5em}
    \caption{Flux at each point observer during the simulation run with larger reference mean free path. The layout is identical to that of Figure~\ref{fig:observer-flux-baseline}.}
    \label{fig:observer-flux-mfp-reference}
\end{figure}

\begin{figure}
    \centering
    \includegraphics[width=1.0\textwidth]{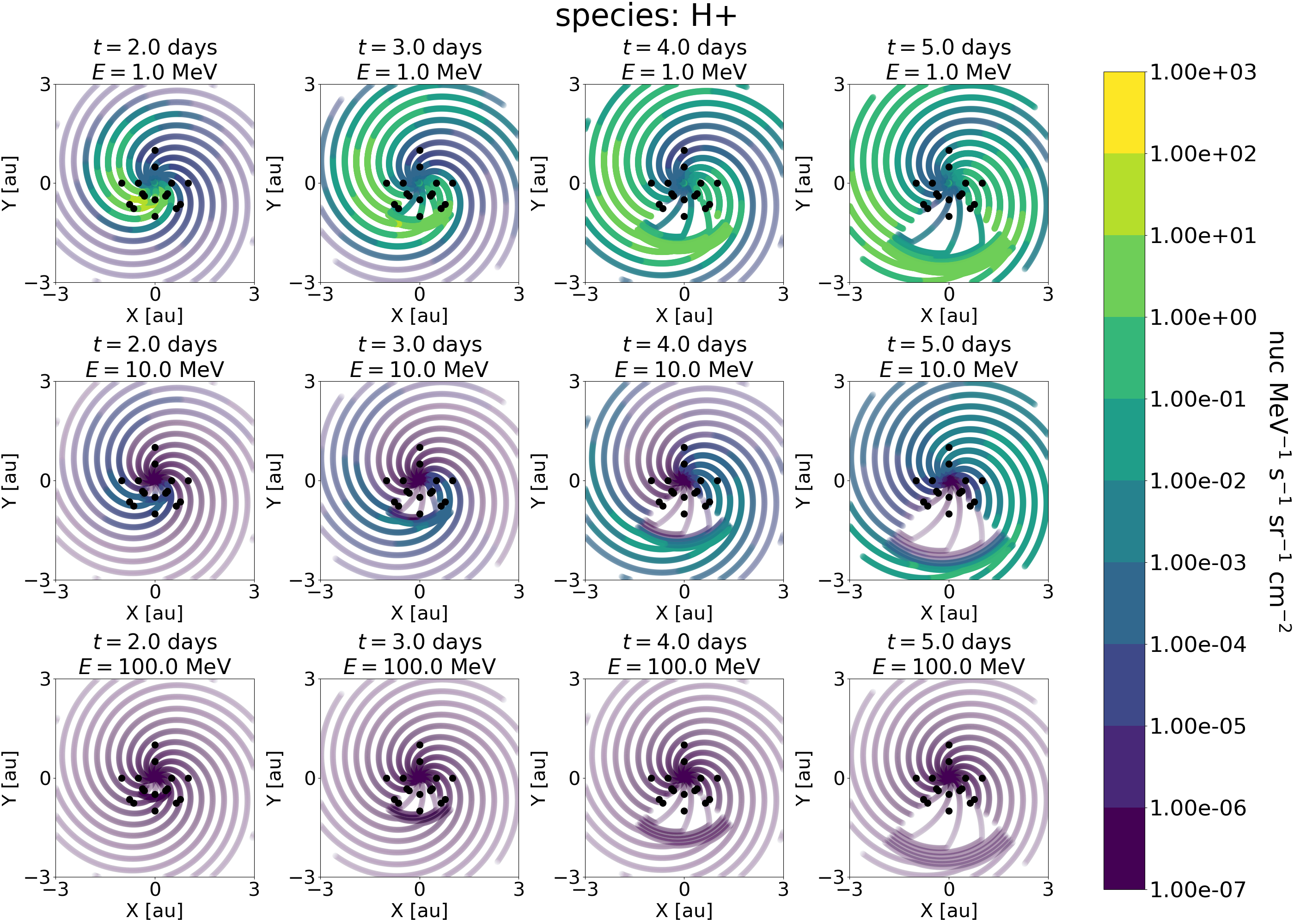}
    \caption{Fluxes of protons during the simulation run with larger reference mean free path. The layout is identical to that of Figure~\ref{fig:stream-flux-baseline}.}
    \label{fig:stream-flux-mfp-reference}
\end{figure}

Figure \ref{fig:observer-flux-mfp-reference} shows proton fluxes from the simulation run with $\lambda_{\parallel 0} = 1.0\unit{au}$. The most noticeable difference from the baseline simulation run is an overall reduction in flux at energies $\geq 1\unit{MeV}$, including the absence of flux at energies $\geq 50\unit{MeV}$. Note the shift to lower energies of fluxes shown in Figure \ref{fig:stream-flux-mfp-reference}. However, increasing $\lambda_{\parallel 0}$ does not simply reduce all fluxes at every observer. Most observers, especially those at 1.0~au, record earlier increases above background in fluxes at 1-10~MeV when compared to the baseline simulation run. The increased reference mean free path also leads to larger peak fluxes of protons with $\leq 1\unit{MeV}$ observed at $-90^\circ$, as well as an earlier peak in protons with those energies at 1.0~au. The observers at $+90^\circ$ record fluxes of protons with 5~MeV and 10~MeV that closely resemble the 50-MeV and 100-MeV fluxes in the baseline run, suggesting that the shapes of these profiles are primarily governed by transport processes that take effect after the protons have been accelerated. Finally, all observers east of the shock origin record more\added{ increased flux during days 4 and 5}, albeit at lower energies, than in the baseline simulation run. \added{The earlier arrival time, larger amplitude, and lower energies are all consistent with an increase to $\lambda_{\parallel 0}$, which not only increases the rate of perpendicular diffusion, but also allows protons to escape from acceleration regions earlier, with correspondingly less energy.}

Figure \ref{fig:stream-flux-mfp-reference} illustrates how 10-MeV protons accelerated by the shock front near 2~au on day 4 propagate along field lines to reach the observers at $180^\circ$ on day 4, then the observers at $-90^\circ$ on day 5. These effects are all consistent with an increase to $\lambda_{\parallel 0}$, which increases the mean free path across the board, allowing protons to escape from acceleration regions earlier, but with less energy, before propagating outward along magnetic field lines. It also allows accelerated protons to propagate farther in a given time frame.

\begin{figure}
    \centering
    \includegraphics[width=1.0\linewidth,trim={2cm 2cm 2cm 2cm},clip=true]{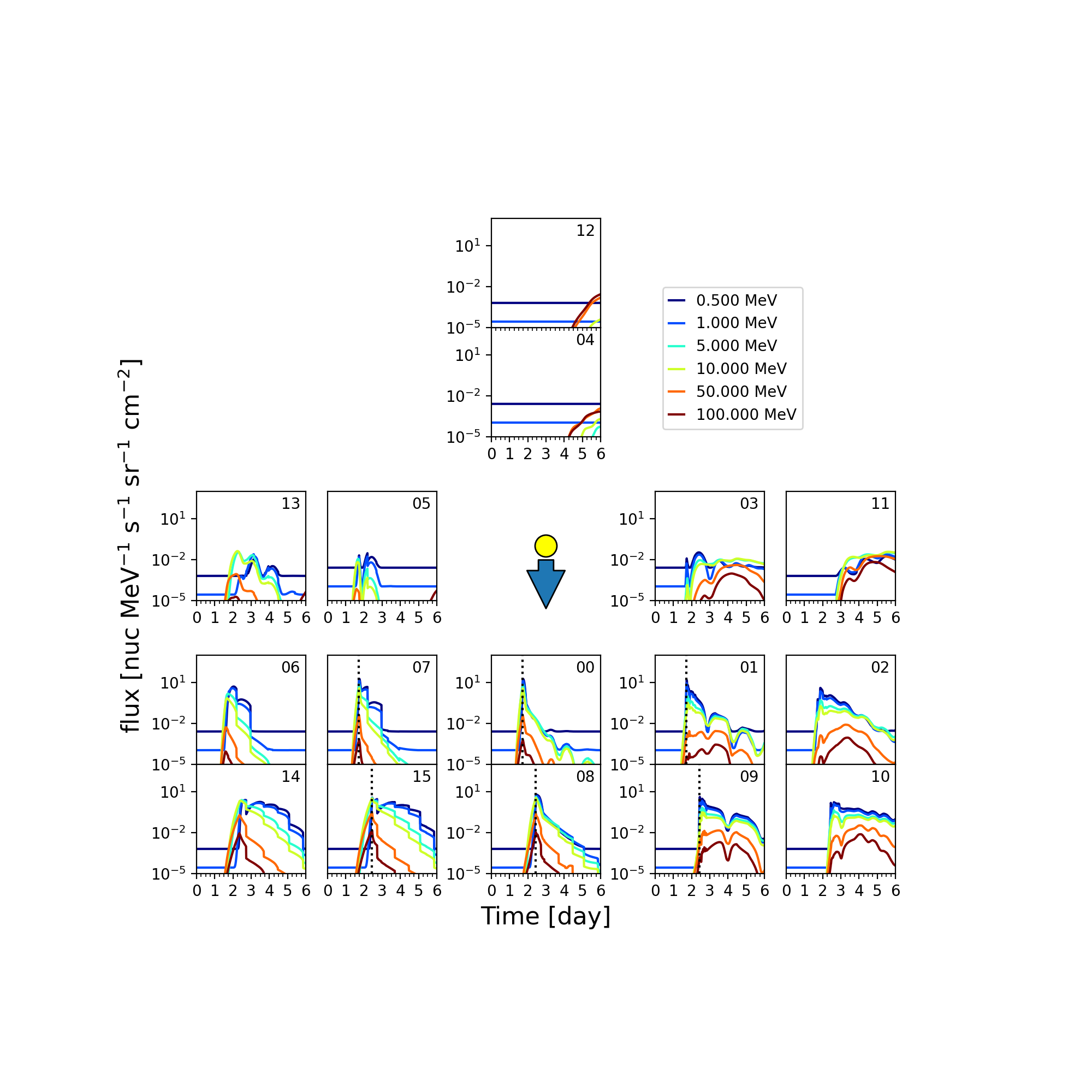}
    \vspace{-5em}
    \caption{Flux at each point observer during the simulation run with higher rigidity power-law index. The layout is identical to that of Figure~\ref{fig:observer-flux-baseline}.}
    \label{fig:observer-flux-mfp-rigidity-hi}
\end{figure}

\begin{figure}
    \centering
    \includegraphics[width=1.0\textwidth]{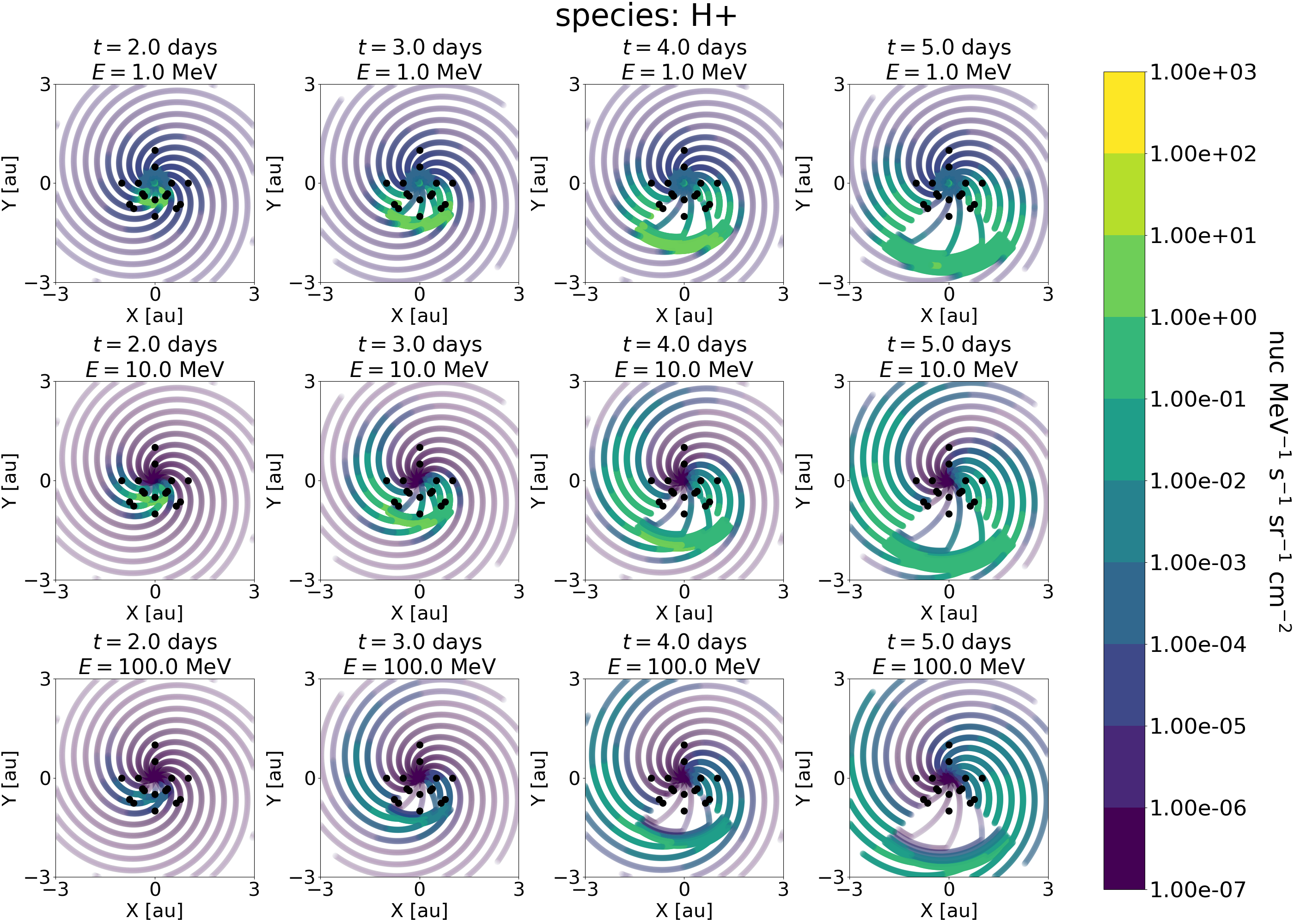}
    \caption{Fluxes of protons during the simulation run with higher rigidity power-law index. The layout is identical to that of Figure~\ref{fig:stream-flux-baseline}.}
    \label{fig:stream-flux-mfp-rigidity-hi}
\end{figure}

Figures \ref{fig:observer-flux-mfp-rigidity-hi} and \ref{fig:stream-flux-mfp-rigidity-hi} show proton fluxes from the simulation run with $\lambda \propto \mathcal{R}^{2/3}$. The overall effect of doubling the power-law dependence of mean free path on rigidity is to reduce the flux of protons with energies $\le 50\unit{MeV}$ and increase the flux of protons with $\geq 50\unit{MeV}$. The flux of 50-MeV protons appears to change very little. Two additional effects worth noting are: 1) the post-shock flux of 500-keV and 1-MeV protons recorded by the 1.0-au observers that straddle the eastern edge of the shock cone exhibits a deeper initial drop-out, compared to baseline values, before recovering to roughly baseline values; and 2) the initial increase in flux of 500-keV and 1-MeV protons recorded by the 1.0-au observer at $-90^\circ$ from the shock origin is significantly delayed in addition to being reduced.

\begin{figure}
    \centering
    \includegraphics[width=1.0\linewidth,trim={2cm 2cm 2cm 2cm},clip=true]{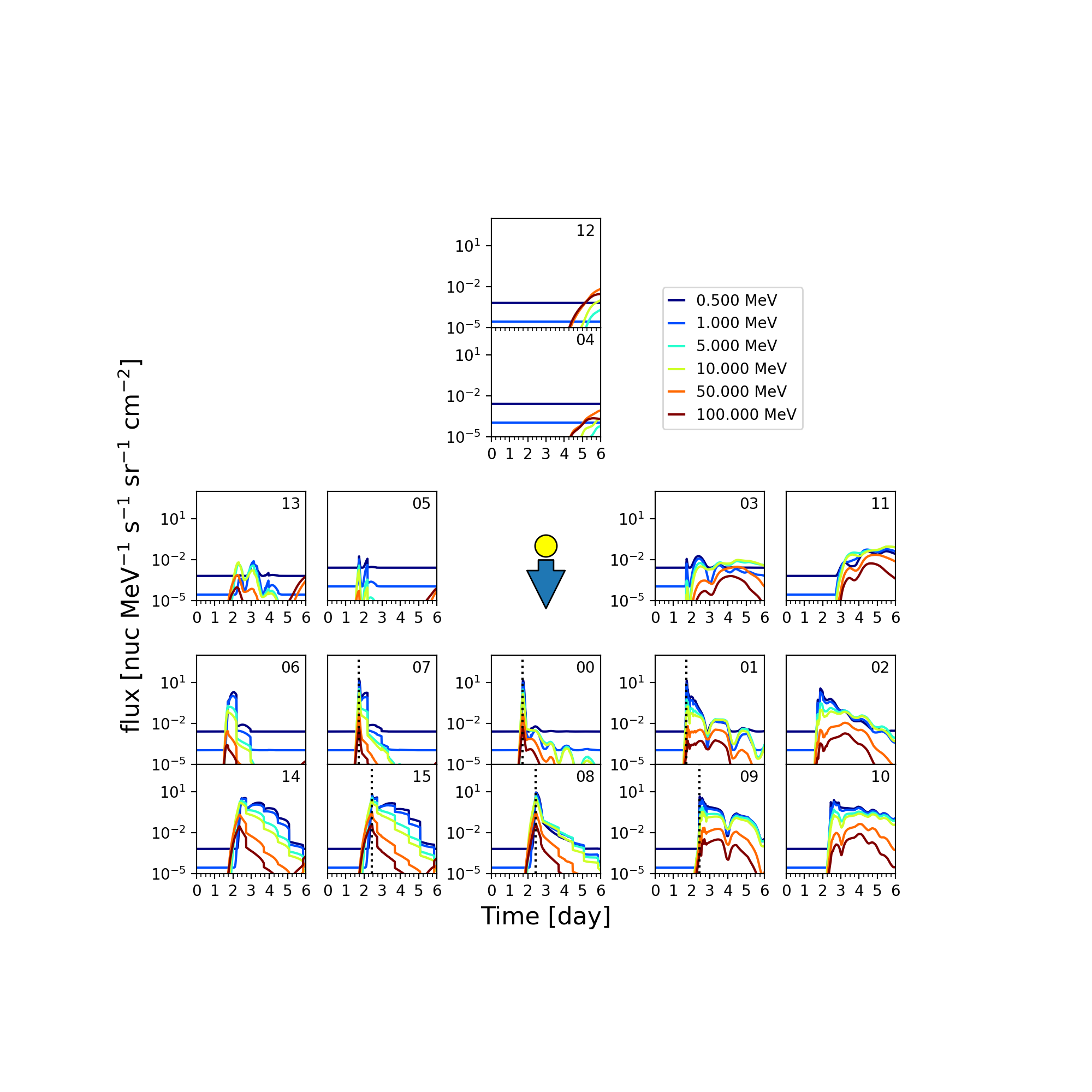}
    \vspace{-5em}
    \caption{Flux at each point observer during the simulation run with higher mean-free-path power-law index (i.e., greater inverse dependence on $|\vec{B}|$). The layout is identical to that of Figure~\ref{fig:observer-flux-baseline}.}
    \label{fig:observer-flux-mfp-power-hi}
\end{figure}

\begin{figure}
    \centering
    \includegraphics[width=1.0\textwidth]{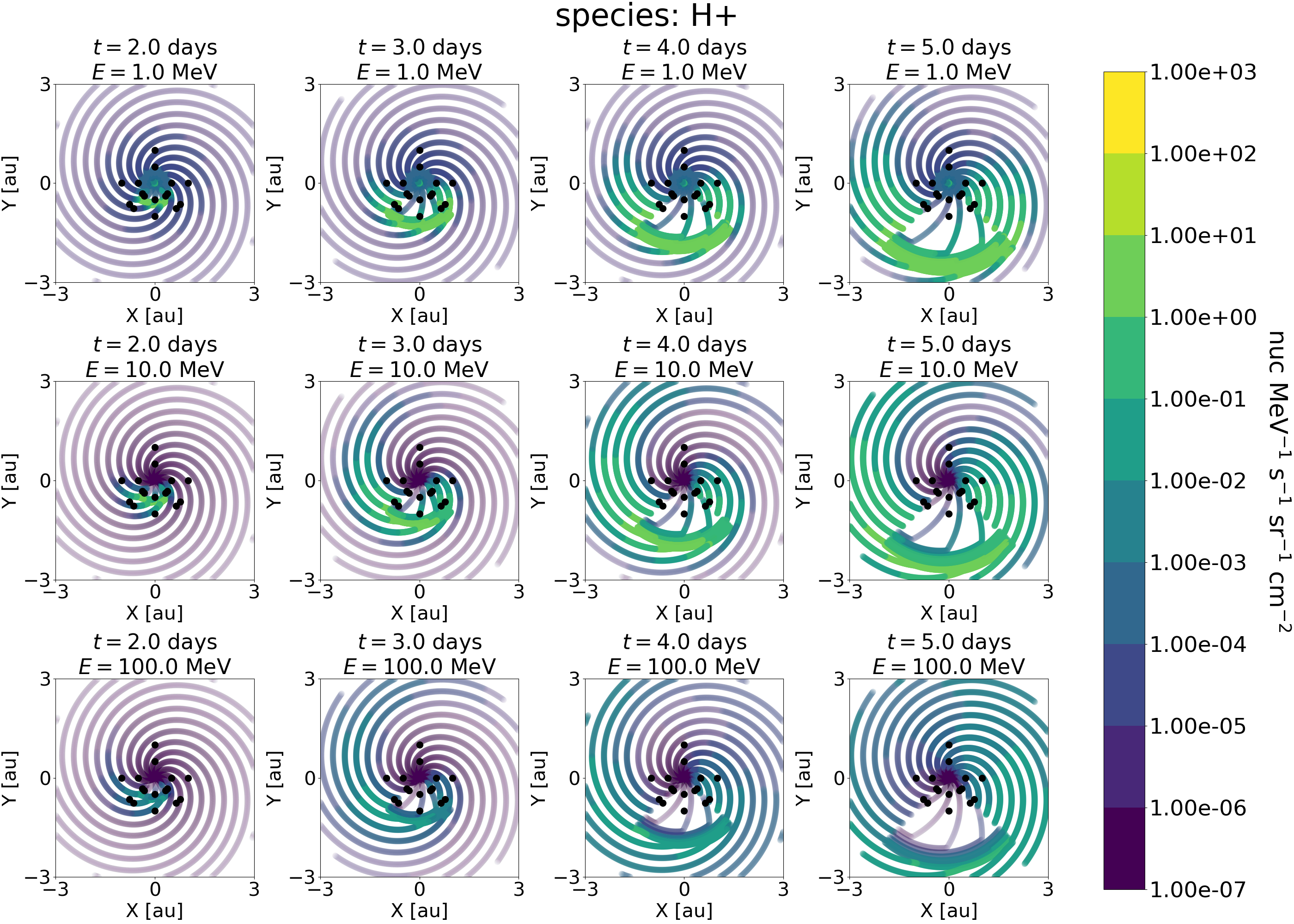}
    \caption{Fluxes of protons during the simulation run with higher mean-free-path power-law index (i.e., greater inverse dependence on $|\vec{B}|$). The layout is identical to that of Figure~\ref{fig:stream-flux-baseline}.}
    \label{fig:stream-flux-mfp-power-hi}
\end{figure}

Figures \ref{fig:observer-flux-mfp-power-hi} and \ref{fig:stream-flux-mfp-power-hi} show proton fluxes from the simulation run with $\beta = 4.0$ (meaning that $\lambda_{\parallel} \propto |\vec{B}|^{-2}$). Overall, this change leads to larger peak fluxes of 100-MeV protons, relative to the baseline simulation run, at most point observers. This increase in high-energy flux accompanies a corresponding decrease in fluxes of protons with $\le 10\unit{MeV}$ at the observers straddling the western edge of the shock cone. The observers located $-90^\circ$ from the shock origin registered the greatest overall reductions in lower-energy fluxes, while the observers located $+90^\circ$ from the shock origin registered noticeable but less significant reductions. The observers centered on and straddling the eastern edge of the shock cone recorded similar peak values of fluxes at $\le 10\unit{MeV}$, accompanied by an increase in higher-energy fluxes, in comparison to the baseline simulation run, but the post-shock falloff is much sharper. Peak fluxes at 1-10 MeV recorded by the observers straddling the western edge of the shock cone differ the least from baseline values, despite the significant increases in higher-energy fluxes, but there is a noticeably faster post-shock reduction, especially at the 0.5-au observer inside the shock cone. The observers at $180^\circ$ registered only modest deviations from baseline fluxes at all energies.

The physical reason for the increased flux of 100-MeV protons corresponding to decreased fluxes of 5-MeV and 10-MeV protons is that increasing the value of $\beta$ when $\lambda_{\parallel} \propto |\vec{B}|^{-\beta/2}$ causes a significant reduction in mean free path in regions where $|\vec{B}|$ is large. This reduced mean free path, in turn, allows for greater acceleration before a proton escapes the local acceleration region \citep{schwadron_particle_2015,young_energetic_2021}. The relative distributions of fluxes in Figure \ref{fig:stream-flux-mfp-power-hi} indicate that less flux at 1 and 10 MeV gets out early in the simulation run while the shock carries a greater flux of 100-MeV protons that diffuse to nearby streams, but not quickly enough to be recorded by the 0.5-au observer inside the western edge of the shock cone.

\begin{figure}
    \centering
    \includegraphics[width=1.0\linewidth,trim={2cm 2cm 2cm 2cm},clip=true]{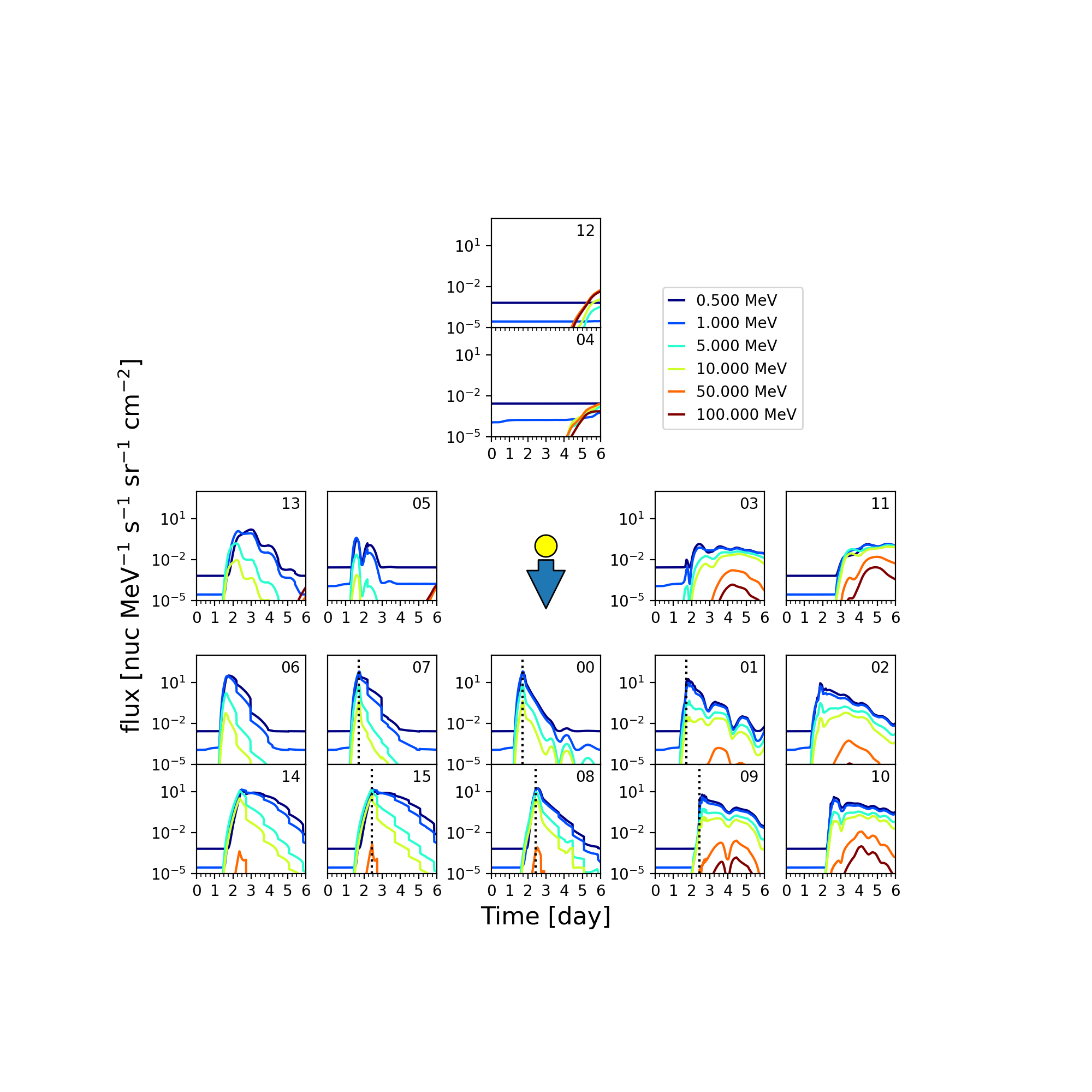}
    \vspace{-5em}
    \caption{Flux at each point observer during the simulation run with lower mean-free-path power-law index (i.e., lesser inverse dependence on $|\vec{B}|$). The layout is identical to that of Figure~\ref{fig:observer-flux-baseline}.}
    \label{fig:observer-flux-mfp-power-lo}
\end{figure}

\begin{figure}
    \centering
    \includegraphics[width=1.0\textwidth]{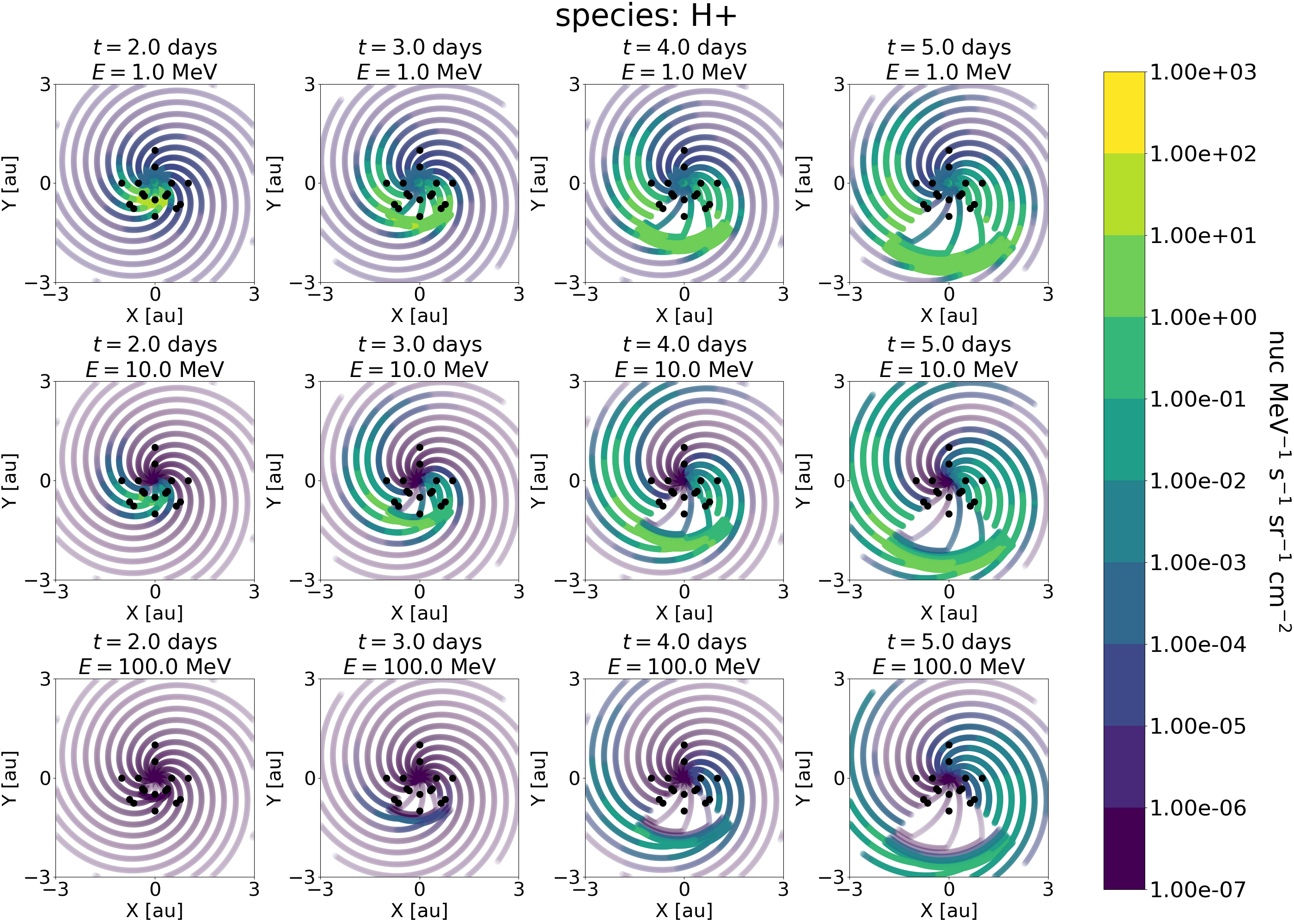}
    \caption{Fluxes of protons during the simulation run with lower mean-free-path power-law index (i.e., lesser inverse dependence on $|\vec{B}|$). The layout is identical to that of Figure~\ref{fig:stream-flux-baseline}.}
    \label{fig:stream-flux-mfp-power-lo}
\end{figure}

Figures \ref{fig:observer-flux-mfp-power-lo} and \ref{fig:stream-flux-mfp-power-lo} show proton fluxes from the simulation run with $\beta = 1.0$ (meaning that $\lambda_{\parallel} \propto |\vec{B}|^{-1/2}$). The reduced dependence on the local magnetic-field strength leads to increased fluxes of protons with 500 keV to 1 MeV and decreased fluxes of protons with higher energies, most notably those with 50 MeV and 100 MeV, at each observer located $0^\circ$ to $-90^\circ$ from the shock origin. All observers at 0.5~au record a gradual ramp up in 1-MeV flux that is not present in the baseline simulation run. Observers straddling the western edge of the shock cone and at $+90^\circ$ from the shock origin record more flux of protons with 50 MeV and 100 MeV, especially the 1.0-au observer $5^\circ$ outside of the shock cone, which records fluxes of 50-MeV and 100-MeV protons during days 3-6 that are nearly equal to their baseline values. However, the signature of the shock front is effectively absent at all locations west of the shock origin in this simulation run. Finally, fluxes recorded by observers at $180^\circ$ from the shock origin differ very little from their baseline values except for a gradual increase in the flux of 1-MeV protons at the 0.5-au observer, starting just before day 1.

These observations are consistent with the inverse sense of the physical reason that increasing $\beta$ when $\lambda_{\parallel} \propto |\vec{B}|^{-\beta/2}$ increased the flux of high-energy protons at the expense of lower-energy protons: A larger mean free path relative to baseline values in regions of enhanced magnetic-field strength allows protons to escape after acceleration to only a few MeV. A fraction of protons that remain in the acceleration regions for long enough to gain energies of 50 MeV or more appear with the shock passage at 1.0-au observers centered on the shock origin and straddling the eastern edge of the shock cone. Another fraction of protons accelerated to at least 50 MeV arrives at the locations of observers west of the shock origin a few days later via the combined effects of perpendicular diffusion and parallel transport along recently connected field lines. The increased flux of protons with $\leq 1\unit{MeV}$ observed east of the shock origin, where protons stream \added{outward along field lines} after acceleration at low radii, likely compensates for the decrease in higher-energy protons observed west of the shock origin.\added{ In other words, those protons were not available to be accelerated by the shock or transported to the western observers}.

\begin{figure}
    \centering
    \includegraphics[width=1.0\linewidth,trim={2cm 2cm 2cm 2cm},clip=true]{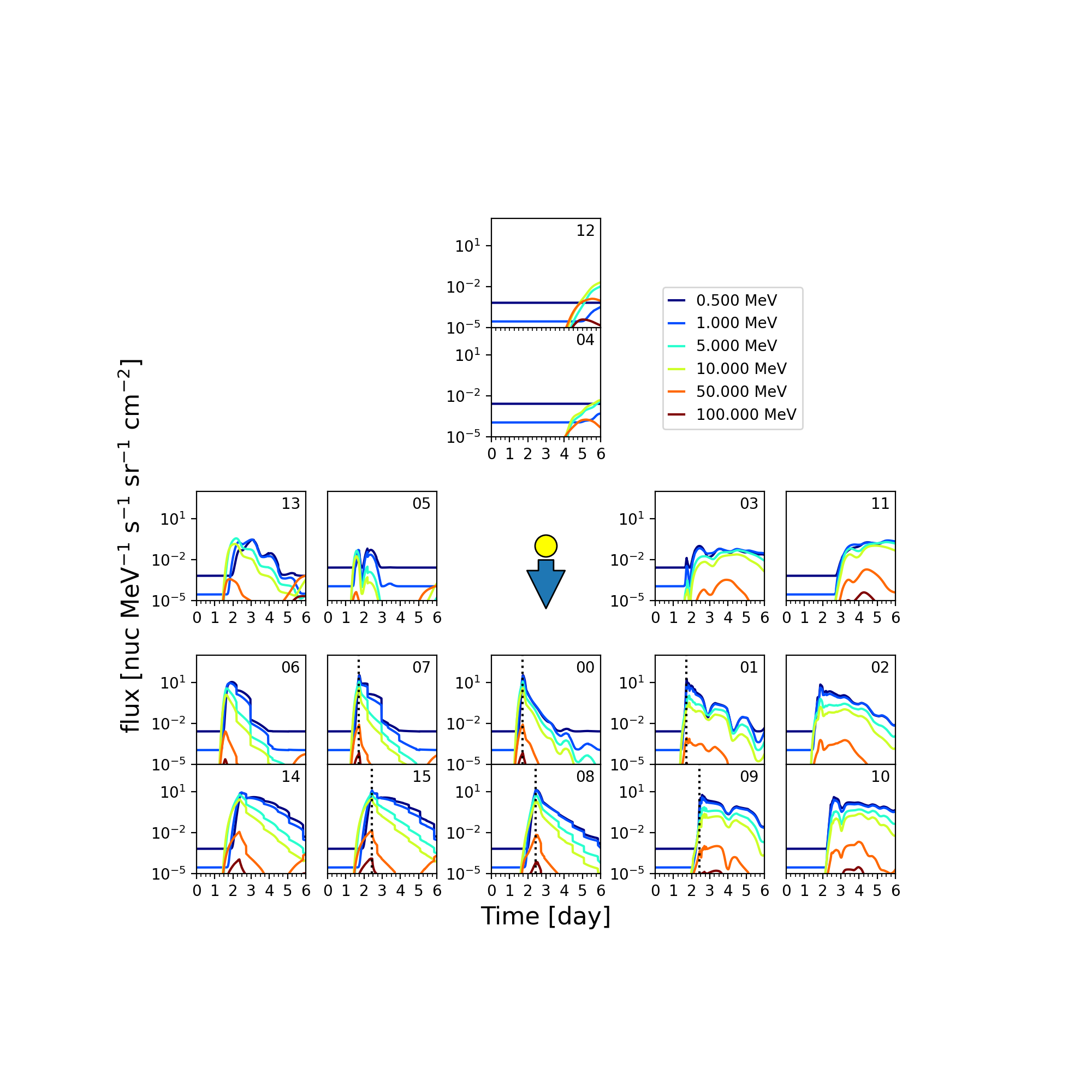}
    \vspace{-5em}
    \caption{Flux at each point observer during the simulation run in which the mean free path increased as the square of the radial distance from the solar surface. The layout is identical to that of Figure~\ref{fig:observer-flux-baseline}.}
    \label{fig:observer-flux-mfp-scaling}
\end{figure}

\begin{figure}
    \centering
    \includegraphics[width=1.0\textwidth]{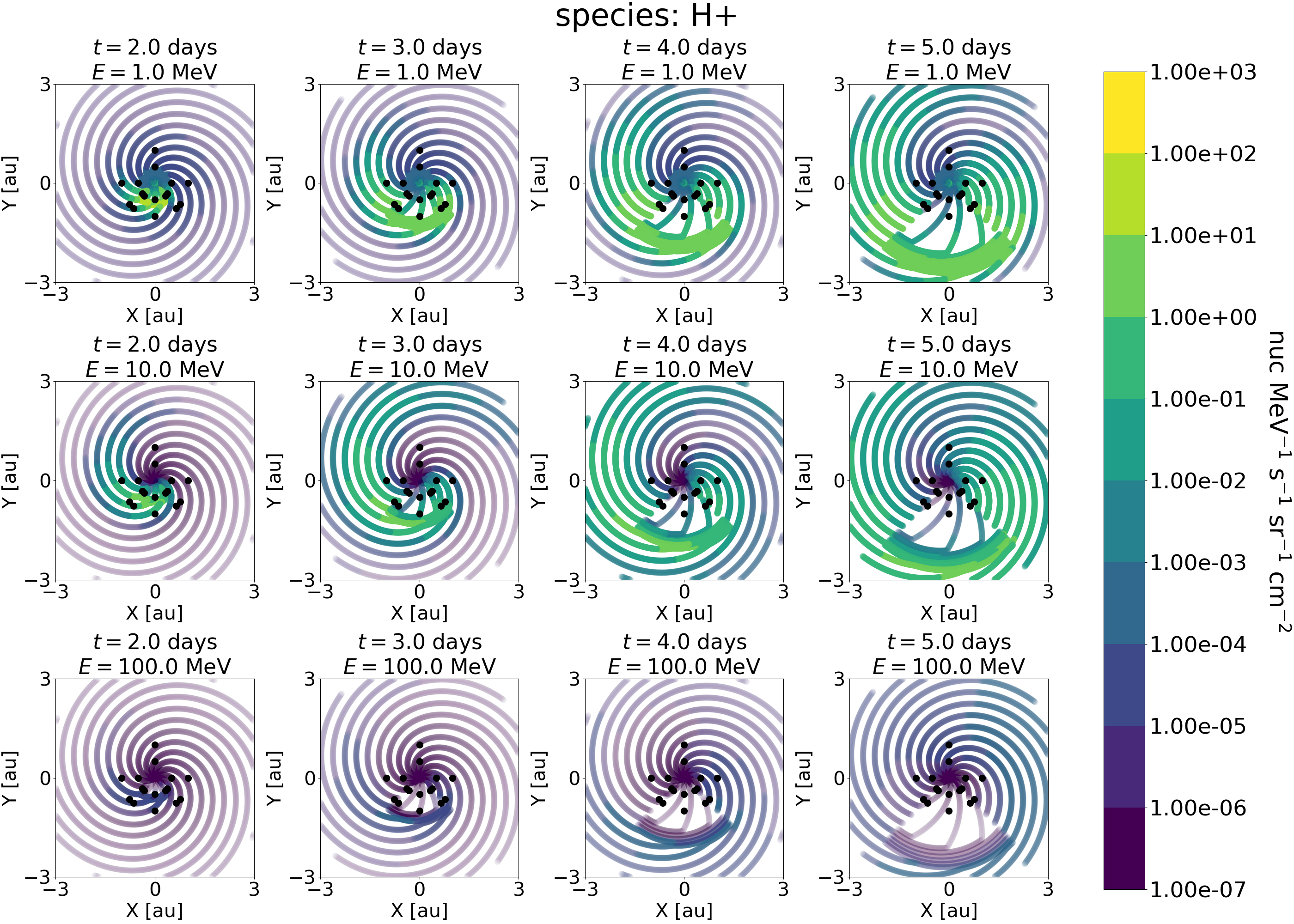}
    \caption{Fluxes of protons during the simulation run in which the mean free path increased as the square of the radial distance from the solar surface. The layout is identical to that of Figure~\ref{fig:stream-flux-baseline}.}
    \label{fig:stream-flux-mfp-scaling}
\end{figure}

Figures \ref{fig:observer-flux-mfp-scaling} and \ref{fig:stream-flux-mfp-scaling} show proton fluxes from the simulation run with $\lambda(r) \propto r^{\beta}$ ($\beta=2$). Fluxes of protons with 1-10 MeV are effectively identical at all observer locations. Higher-energy fluxes at the $-90^\circ$ observers are nearly identical to baseline values up to\added{ the end of} day 4, at which point \added{fluxes of 50-MeV }protons reach these observers. This onset is 12-24 hours earlier than in the baseline run, depending on energy. This behavior extends to the 0.5-au observers straddling the eastern edge of the shock cone, with 1.0-au observers recording lower peak values at 50 and 100~MeV, in addition to an earlier onset of protons\added{ tranported via perpendicular diffusion}. Observers west of the shock origin generally record lower fluxes at 50 and 100~MeV, with steeper post-shock decreases. The observers at $+90^\circ$ also record slightly higher post-shock 1-MeV fluxes when compared to baseline values. The observers at $180^\circ$ from the shock origin record larger fluxes at energies up to 10~MeV, again corresponding to a reduction in fluxes at higher energies, with an increase at 1~MeV during day 5 that is absent in the baseline simulation run. Furthermore, the increased flux at 50 and 100~MeV peaks during day 5 and is decreasing by the end of the simulation run, while lower-energy fluxes appear to be increasing. Figure \ref{fig:stream-flux-mfp-scaling} confirms that, though \added{shock-accelerated protons with energies up to 100~MeV reach eastern observers} early in the simulation run, most of the 100-MeV flux \added{reaches observers outside the shock cone days later via a combination of parallel and perpendicular diffusion}.

\subsection{The Role of Shock Profile}
\label{sec:The Role of Shock Parameters}

\begin{figure}
    \centering
    \includegraphics[width=0.75\linewidth]{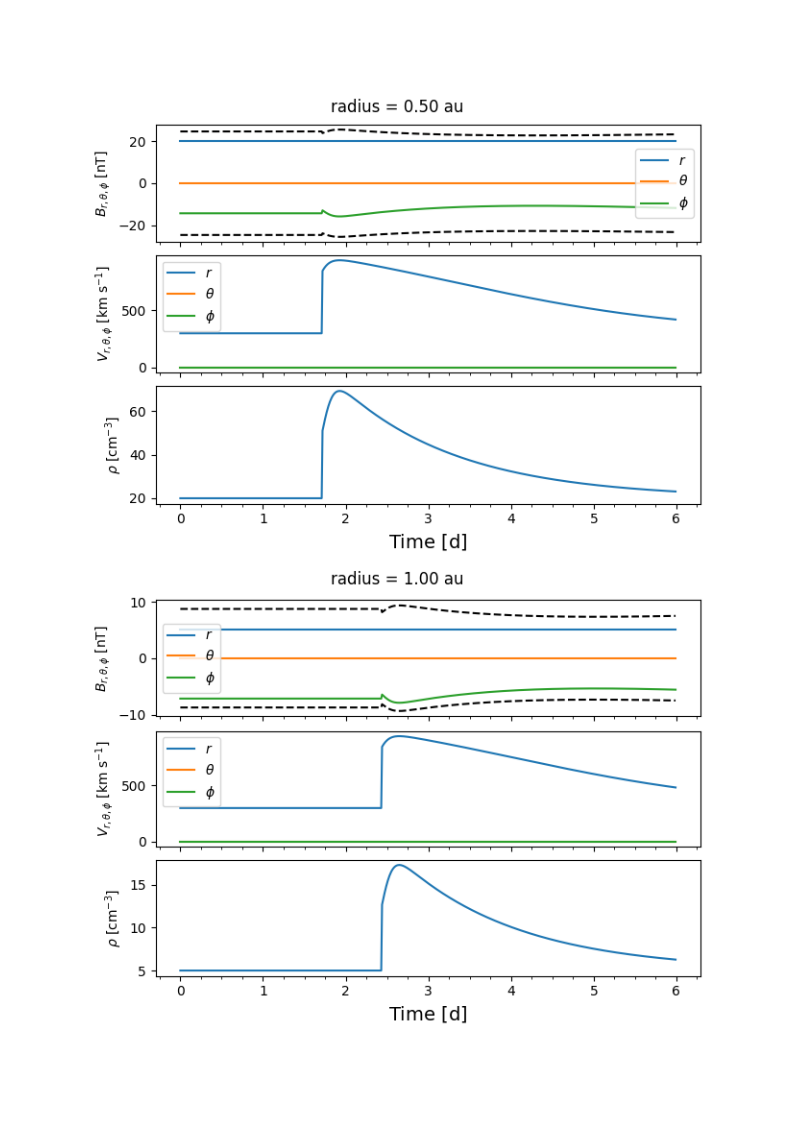}
    \vspace{-5em}
    \caption{MHD quantities from a simulation run with $\Gamma_{s} = 10$, using the same layout as Figure \ref{fig:observer-mhd-baseline}.}
    \label{fig:observer-mhd-shock-profile}
\end{figure}

To examine the role of the shock profile in accelerating protons, we ran a variation of the baseline simulation using a more gradual shock front by setting $\Gamma_{s} = 10$. The EPREM parameter \texttt{idealShockGradient}, which corresponds to $\Gamma_{s}$, controls the deviation of the ideal-shock front from a step function. Reducing $\Gamma_{s}$ from 100 to 10 results in an ideal shock with a more gradual transition. Figure \ref{fig:observer-mhd-shock-profile} shows the MHD quantities as functions of time with this modified shock profile. In comparison to Figure \ref{fig:observer-mhd-baseline}, the peak density, radial velocity component, and azimuthal magnetic-field component all exhibit a more gradual increase to their peak after shock arrival.

\begin{figure}
    \centering
    \includegraphics[width=1.0\linewidth,trim={2cm 2cm 2cm 2cm},clip=true]{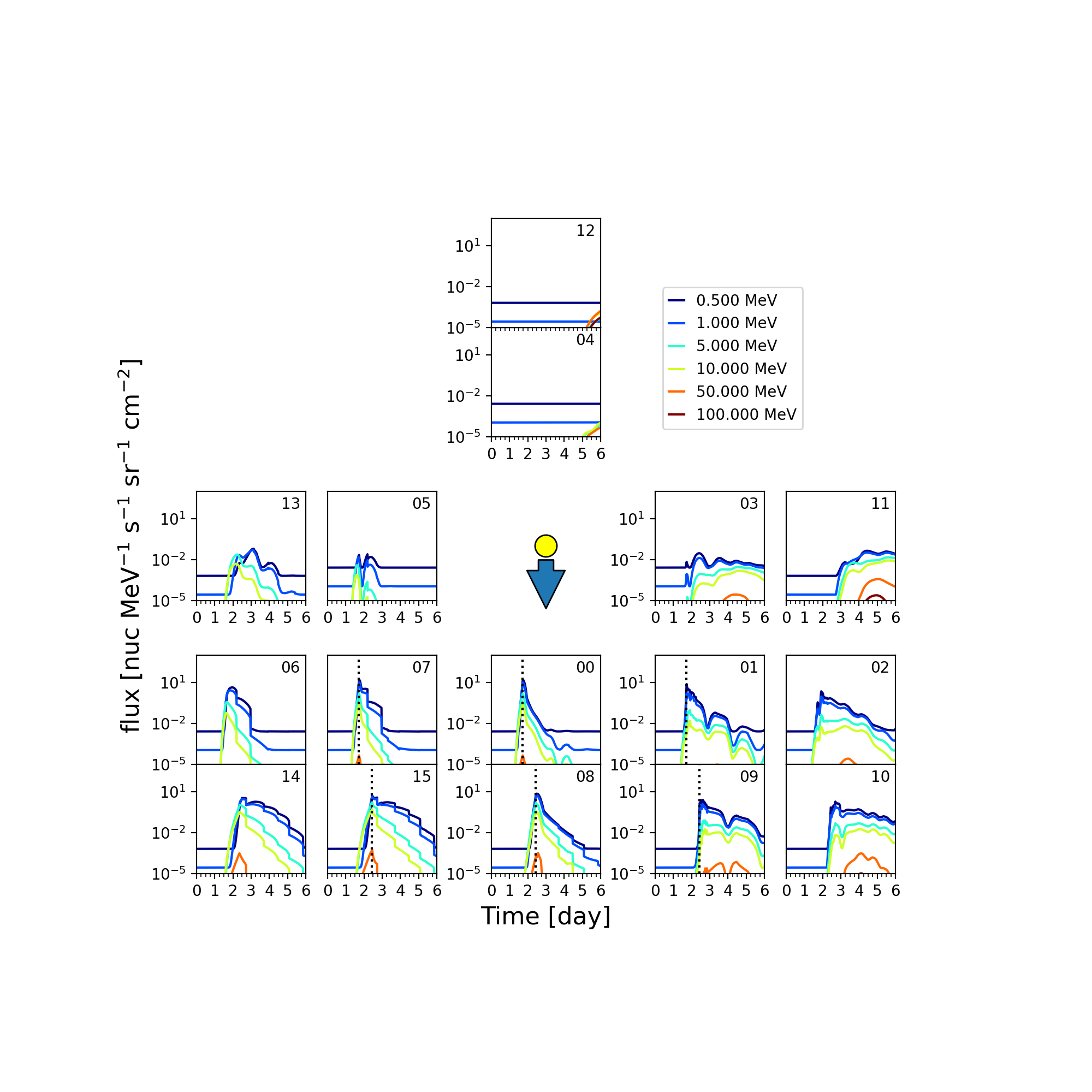}
    \vspace{-5em}
    \caption{Flux at each point observer during the simulation run in which the ideal-shock gradient was more gradual. The layout is identical to that of Figure~\ref{fig:observer-flux-baseline}.}
    \label{fig:observer-flux-shock-profile}
\end{figure}

\begin{figure}
    \centering
    \includegraphics[width=1.0\textwidth]{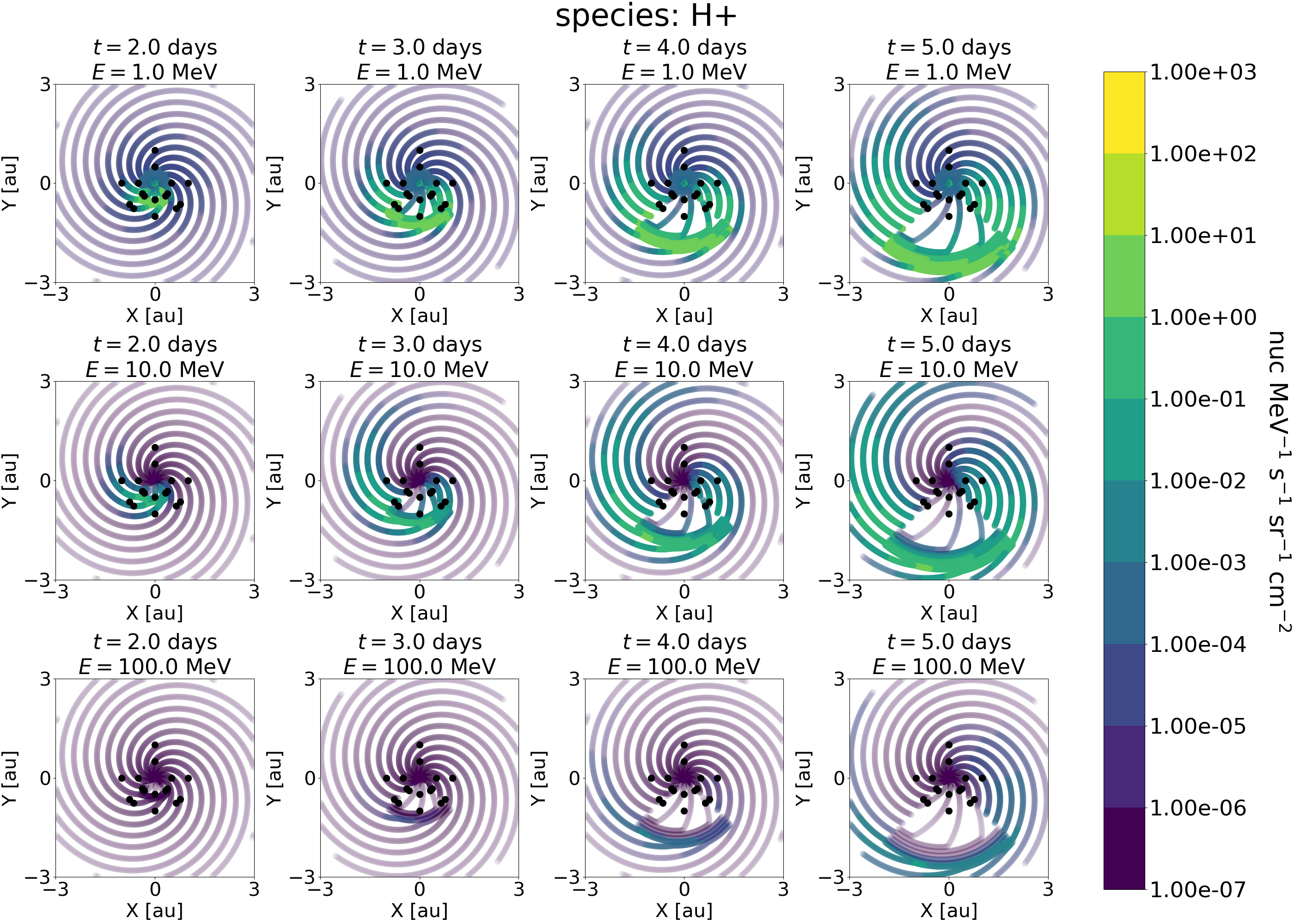}
    \caption{Fluxes of protons during the simulation run in which ideal-shock gradient was more gradual. The layout is identical to that of Figure~\ref{fig:stream-flux-baseline}.}
    \label{fig:stream-flux-shock-profile}
\end{figure}

Figures \ref{fig:observer-flux-shock-profile} and \ref{fig:stream-flux-shock-profile} show proton fluxes from the simulation run with $\Gamma_{s} = 10$. The major effect of softening the shock-front gradient is to reduce fluxes below their baseline values at all observers, with the amount of reduction increasing with energy. This broad reduction resulted in the observers located $180^\circ$ from the shock origin recording only negligible flux in the final day of simulation time. The softer shock front also causes fluxes observed at locations $0^\circ$--$90^\circ$ from the shock origin to return to their pre-shock values more quickly. West of the shock origin, the flux of protons with $\leq 10\unit{MeV}$ is more distinctly separated into a band at 500~keV--1~MeV, and one at 5~MeV--10~MeV.

\added{A reasonable explanation for the reduced fluxes shock in Figures \ref{fig:observer-flux-shock-profile} and \ref{fig:stream-flux-shock-profile} is that the softer shock gradient reduces the effective compression ratio that protons encounter as they pass through the shock front. Using the notation $\{Q\}$ to represent the ratio of MHD quantity $Q$ in the ``shock-profile'' simulation run versus that of the baseline simulation run, $\min\{B_{\phi}\} = 0.79$, $\min\{V_{r}\} = 0.90$, and $\min\{\rho\} = 0.71$. In other words, the shocked values of $B_{\phi}$, $V_{r}$, and $\rho$ were respectively 21\%, 10\%, and 29\% weaker in the ``shock-profile'' simulation run (with $\Gamma_{s} = 10$) than in the baseline simulation run (with $\Gamma_{s} = 100$).}

\section{Discussion}
\label{sec:Discussion}

Despite EPREM's simple built-in shock model, the simulation runs presented in \S \ref{sec:Simulation Results} can tell us a good deal about the nature of widespread SEP events. First of all, it is not difficult to produce flux enhancements at energies up to 100 MeV over a span of $\pm 90^\circ$ from the shock origin, though the ability to observe these flux enhancements depends not only on an observer's location but also on the sensitivity of a given instrument. Assuming a typical noise floor of $10^{-3}\fluxunit$ \citep[e.g.,][]{kollhoff_first_2021,dresing_17_2023}, most simulation runs shown here generated observable flux of 100-MeV protons at the location of at least one observer. Of those that caused one or more observers to not observe 100-MeV protons above this fiducial threshold, increasing the reference parallel mean free path, $\lambda_{\parallel 0}$, played the most significant role (cf. Figure \ref{fig:observer-flux-mfp-reference}), followed by weakening the shock gradient (cf. Figure \ref{fig:observer-flux-shock-profile}). Setting $\lambda_{\parallel}\left(r\right) \propto r^{-2}$ everywhere (cf. Figure \ref{fig:observer-flux-mfp-scaling}) preferentially reduced 100-MeV fluxes west of the shock origin while setting $\lambda_{\parallel}\left(|\vec{B}|\right) \propto |\vec{B}|^{-1/2}$ preferentially reduced 100-MeV fluxes at and east of the shock origin.

We can test the general validity of our baseline simulation run by comparing the fluxes shown in Figure \ref{fig:observer-flux-baseline} to those shown in the \added{right} panel of Figure~1 in \citet{dresing_17_2023}. During the 17~April 2021 widespread SEP event, BepiColombo was best connected to the flare site, at a radial distance of 0.63~au, and therefore best corresponds to our point observer~00. It observed a sharp increase in the flux of 25.1~MeV protons roughly one hour after the associated flaring event, peaking at approximately $4-5\times 10^{-1}\fluxunit$, then exponentially decaying to the background level over the following 30~hours. The baseline flux of 10-MeV protons recorded by observer 00 qualitatively mirrors this behavior; in fact, assuming the peak flux of 25.1~MeV protons should lie between those of 10~MeV and 50~MeV protons,\added{ it} quantitatively agrees reasonably well. Point observer 07 recorded a slightly lower peak flux and faster decay, similar to that observed by Parker Solar Probe, while point observers 02 and 03 recorded flatter profiles with still lower peaks, similar to \added{fluxes at 19.65 to 25.09~MeV} observed by Solar Orbiter. There is even some consistency between point observer~11 and STEREO-A, which observed weak fluxes of protons with 20.8 to 23.8~MeV, beginning later than the peak intensity observed by BepiColombo and Parker Solar Probe.

\added{One clear discrepancy between the baseline simulation run and the observational catalog is the continual generation of protons with $\geq 50$ MeV over the course of multiple days, as recorded by observers west of the ideal shock cone (e.g., observer 11). The most likely explanation is that the ideal shock propagates outward with a constant radial speed of 1200 km/s. While many CME-driven shocks in the solar wind originate around 2–3 solar radii with speeds $\sim 2000\unit{km/s}$, such high shock speeds are rarely observed beyond a few tens of solar radii. In reality, CMEs typically undergo a period of rapid deceleration following their initial acceleration phase, after which the CME propagates with nearly constant speed \citep[cf.][]{reiner_coronal_2005,liu_sun--earth_2013}.\added{ For example, \citet{cohen_pspisis_2021} showed that the CME that produced the 29 November 2020 widespread SEP event originated with a speed of nearly 2000 km/s before decelerating to approximately 1500 km/s by the time it reached 25 \Rsun.} The ideal shock model within EPREM as of v0.14.0 is incapable of capturing any of these dynamics. However, the ability to specify periods of acceleration and deceleration are in development and will be the subject of future studies.}

In terms of observing a widespread event, two cases significantly reduced overall flux recorded by the observer at $180^\circ$ from the shock origin: the simulation run without perpendicular diffusion (cf. Figure~\ref{fig:observer-flux-no-perp-diff}) and the simulation run with the weaker shock (cf. Figure~\ref{fig:observer-flux-shock-profile}). It is, perhaps, not surprising that a weaker ideal shock reduced fluxes of protons with 50 and 100~MeV, and also reduced overall flux at the observer farthest from the shock origin, but we must note that it still produced substantial flux of protons with energies up to 10~MeV at most observers. Conversely, increasing $\lambda_{\parallel 0}$ from 0.1~au to 1.0~au produced the highest overall flux at the observer farthest from the shock origin. That flux was dominated by protons with energies up to 5~MeV, but it would almost certainly have been observable. The simulation run in which $\lambda_{\parallel}\left(r\right) \propto r^{-2}$ also produced slightly higher fluxes of protons with 5 and 10 MeV at the farthest observer, in comparison to baseline values, but the increase is not nearly as drastic.

The mean-free-path inverse dependence on $|\vec{B}|$ also affects diffusion. On one hand, streams from run mfp-power-hi show larger fluxes of 100-MeV protons near $180^\circ$ with broader radial spread. This is consistent with the mean free path growing even more quickly with radial distance, which allows these protons to diffuse outward more quickly, and can thus account for the earlier reduction in 100-MeV protons recorded by observer 04 (at $180^\circ$ and 0.5~au). On the other hand, observer 04 recorded a slight increase in lower-energy protons during run mfp-power-lo, for which the mean free path had the least dependence on $|\vec{B}|$ of all simulation runs shown here. This reduced dependence on $|\vec{B}|$ leads to reduced perpendicular diffusion at all energies, compared to baseline values. However, it also reduces the maximum energy that protons acquire before leaving the acceleration region \citep{schwadron_particle_2015}. We can therefore attribute the slight increase in 1-MeV protons recorded by observer~04 to diffusion from regions near observers~05 and 13 (i.e., far east of the shock origin).

Changing the dependence of mean free path on magnetic-field strength and rigidity had the least noticeable effect on fluxes observed at $180^\circ$ from the shock origin, especially for protons with 50 and 100~MeV at 1.0~au. However, the values of those parameters strongly affected timing of peak fluxes and the evolution of flux as a function of energy at all other observers. This means that simulation runs designed to reproduce proton fluxes observed by spacecraft may need to tune the value of each parameter in order to improve agreement. Quasi-linear theory developed by \citet{jokipii_cosmic-ray_1966,jokipii_addendum_1968} and extended by \citet{erdos_scattering_1999} with the help of Ulysses observations suggests $\lambda_{\parallel}\left(B\right) \propto |\vec{B}|^{-\beta/2}$ with $2/3 \leq \beta \leq 2$. The collection of simulation runs presented here include two values of $\beta$ within that range (namely, 1 and 2) and one value of $\beta$ above the proposed upper bound (namely, 4). The latter value had a drastic effect on the simulated fluxes recorded by point observers, leading to possibly non-physical results, but those results are nonetheless valuable if they can explain unusual \emph{in situ} observations by spacecraft. Fluxes in a simulation run with $\beta = 2/3$ (not shown) continue two trends seen in the mfp-power-hi, baseline, and mfp-power-lo runs: further decreasing $\beta$ leads to 1) further reduced fluxes at energies $> 1\unit{MeV}$ and 2) a stronger pre-shock ramp-up in fluxes at energies $\leq 1\unit{MeV}$. \citet{droge_rigidity_2000} derived estimates of the mean free path dependence on rigidity by computing fits to the ISEE-3/ICE spacecraft. They found a consistent power-law index of roughly 0.3 for ions despite large variation in mean-free-path amplitude, which ranged from 0.02~au to more that 1.0~au. EPREM\added{ v0.14.0 is not capable} of running with $\chi = 0$ (i.e., a mean free path that is constant with respect to rigidity) because the lower bound is set at $10^{-33}$. However, fluxes from a run with $\chi = 10^{-33}$ (not shown) continued the main trends between $\chi = 2/3$ in the mfp-rigidity run to $\chi = 1/3$ in the baseline run: higher flux of protons with energy $\leq 1\unit{MeV}$, lower flux of protons with energy $\geq 50\unit{MeV}$, and flux increases that occur essentially contemporaneously across energies east of the shock origin.

\citet{dresing_17_2023} note that PSP and SolO, which had similar longitudinal separations (east and west, respectively) from the flaring active region on 17 April 2021, observed dramatically different SEP characteristics. As they note, PSP observed a more intense and impulsive event while Solar Orbiter observed a more gradual, less intense, and delayed event. These observations are consistent with overall trends in fluxes recorded by simulated point observers shown here: Observers straddling the eastern edge of the ideal shock cone recorded more impulsive flux profiles, while those straddling the western edge recorded more gradual flux profiles, especially at energies $\geq 10$ MeV. Whereas \citet{dresing_17_2023} state that the difference in flux profiles suggests differential magnetic connection to the source region due possibly to connection to different portions of the shock (cf. the canonical description of \citet{cane_role_1988}), to a combination of differently directed SEP injections, or a mixture of those two factors, the difference in our simulated flux profiles can only be due to differences in connectivity.

We may also make our own direct comparison to the canonical flux profiles published in Figure 15 of \citet{cane_role_1988}, as well as in Figure~2 of \citet{reames_how_2023}. The baseline profiles of flux at 50-100 MeV recorded by observers at $0^\circ$ from the shock origin and straddling the western edge of the shock cone exhibit features similar to path D in Figure~15 of \citet{cane_role_1988}: observers at 0.5~au see more of the sharp initial increase while observers at 1.0~au see the initially sharp, then more gradual, decrease. The slightly more gradual initial increase recorded by observers at 1.0~au during the baseline run more closely mimics panels~(c) and (d) in Figure~2 of \citet{reames_how_2023}, and Figure~\ref{fig:observer-flux-mfp-reference} shows a much more gradual pre-peak increase, suggesting the presence of a larger reference mean free path upstream of the acceleration region during the observed events. Simulated fluxes of protons with energy $\leq 10$~MeV recorded by the observer at 1.0~au inside the eastern edge of the shock cone during the baseline run exhibits increasing and decreasing periods that are similar to those in path A of \citet{cane_role_1988}, including a significant dip after a few days. Again, increasing the simulation's reference mean free path leads to broad similarity with observed fluxes, this time between observer 09 in Figure~\ref{fig:observer-flux-mfp-reference} and panel~(f) of \citet{reames_how_2023}. The simulated flux of 50-MeV protons tends to follow the observed flux in path C of \citet{cane_role_1988}, though not necessarily panel (g) of \citet{reames_how_2023}, except for the simulation run with larger reference mean free path, in which the fluxes of protons with 5 and 10 MeV better match the composite observations shown by \citet{cane_role_1988}. The best agreement between panel~(a) of \citet{reames_how_2023}, for which there is no corresponding path in the figure from \citet{cane_role_1988}, appears to come from the simulation run with lower mean-free-path power-law dependence on $|\vec{B}|$ (Figure~\ref{fig:observer-flux-mfp-power-lo}). This is consistent with acceleration at a quasi-parallel shock, where the density compression primarily does the work on accelerating ions to SEP energies.

A major difference in all these comparisons is in the time of shock arrival relative to the flux peak; we attribute this to the fact that the CMEs that produced the flux profiles shown by \citet{cane_role_1988} and \citet{reames_how_2023} tended to slow down before reaching IMP~8 at 1.0~au whereas our idealized cone shock propagates radially outward at a constant 1200~km/s.\added{ As previously stated, development of more realistic ideal-shock acceleration and deceleration is in progress, and will be the subject of future studies.} Another significant difference is that our idealized cone shock does not include realistic compression regions that can sweep up particles, which spacecraft on the western flank will observe.\added{ This is related to the difference in geometry between the simulated cone shock and physical CME-driven shocks. A CME expanding into the solar corona will produce a quasi-parallel shock along its eastern flank and a quasi-perpendicular shock along its western flank, due to the orientation of the CME front relative to the spiral magnetic field (assuming a single-CME scenario). As the EPREM ideal-shock cone expands into the simulated solar wind, it encounters upstream nodes in a predominantly quasi-parallel sense along its entire front until about 1~au, where the spiral angle goes through $45^\circ$.}

\added{An important related caveat is that the shock-cone's radial propagation causes the transition from quasi-parallel to quasi-perpendicular geometry to first occur at its \emph{eastern} flank before traveling westward, whereas the more field-aligned propagation of a realistic CME-driven shock would favor quasi-parallel geometry on its eastern side. At small radial distances, magnetic field lines are mostly radial, so the expanding cone should produce a quasi-parallel shock that is overall consistent with a CME-driven shock. Upstream simulated flux profiles should therefore represent physically meaningful results. Near and downstream of the ideal shock, simulated proton fluxes recorded by point observers in or near the shock cone may deviate from observations. These discrepancies are further exacerbated by the fact that the downstream region does not currently include any model of flux ropes, cavity structures, or enhanced turbulence.}

Finally, \citet{marsh_drift-induced_2013} argued that drifts can significantly alter the propagation of SEPs with large enough energies. A simulation run with perpendicular drift (not shown) produced fluxes that were at most $\approx 1.7$~times larger than the corresponding baseline values, with greater increases indeed occurring in the high-energy bins, and the relative increases all occurring after shock passage. Otherwise, there was effectively no difference from the baseline simulation run. The reason that perpendicular drifts do not seem to have a pronounced effect in our simulated SEP event could be that our simulation runs included only protons, whereas \citet{marsh_drift-induced_2013} found that the effect is stronger for ions with larger mass-to-charge ratios, or because our simulation runs spanned too short a time to notice the effect.

\section{Concluding Remarks}
\label{sec:Concluding Remarks}

EPREM is an ever maturing model of ion acceleration and transport that solves the focused transport equation on a Lagrangian grid of nodes, given time-dependent values of the magnetic field, velocity, and density in the solar wind. Alternative implementations (many of which are no longer in use) include one-way coupling to a specific MHD model that provides the requisite time-dependent MHD quantities. The implementation described here internally generates those quantities in the form of a idealized cone shock. A primary task in the development roadmap for EPREM will be to add one-way MHD coupling that is agnostic to the origin of the MHD data. The appendices to this article provide details of key features and elements of EPREM.

The current era of spacecraft exploration, in which it is possible to simultaneously observe heliospheric phenomena over large separations in radius and longitude, provides excellent opportunities to test global heliospheric models like EPREM. In particular, observations of widespread SEP events as reported in the literature are apt test scenarios against which to compare the results of numerical simulations that aim to model SEP fluxes at multiple points over time. Inversely, the results of those simulations can help the heliophysics community better understand the driving factors behind observed SEP dynamics.

This work has presented 8 variations of a simulated SEP event using EPREM with a simple cone-shock model. Despite the simplicity of the shock, all simulation runs exhibited complex profiles of SEP flux as a function of time and energy, with clear dependence on parameters related to diffusion, mean free path, and shock profile. All simulation runs also exhibited significant longitudinal spread in SEP flux, but certain parameter values led to a decrease or absence in SEP flux at observers located $\geq 90^\circ$ from the shock origin. Understanding the specific differences in SEP flux caused by changing the value of each parameter in the simulation runs presented here can provide insight into the morphology of observed CME-driven SEP events and the state of the solar wind through which the CME propagates. These results are especially timely for investigating the origin widespread SEP events and for predicting SEP fluxes at key locations throughout the heliosphere.

The code use to create the figures shown in this work consists of a set of Bash scripts that call routines from the open-source package \url{https://gitlab.com/open-eprem/eprem-analysis}, which is based on the open-source package \url{https://gitlab.com/open-eprem/eprempy}. The Bash scripts, as well as the source code for EPREM v0.14.0, are publicly available in \citet{young_2025_17195904}\added{ [on Zenodo:\dataset[doi:10.5281/zenodo.17195904]{https://doi.org/10.5281/zenodo.17195904}]}. EPREM v0.14.0 is also available as a release tag in the git repository at  \url{https://gitlab.com/open-eprem/eprem}. The runtime-configuration (input) and point-observer (output) files from each simulation run performed for this project are publicly available in \citet{young_2025_17196185}\added{ [on Zenodo:\dataset[doi:10.5281/zenodo.17196185]{https://doi.org/10.5281/zenodo.17196185}]}. The full collection of simulation output files, which comprise 2.5 TB, requires an alternative storage solution and is not currently available in a public archive. Interested readers are encouraged to contact the authors to learn if the data have since become publicly available or to discuss access options. The authors are committed to making the data as accessible as possible.

\begin{acknowledgements}
This work is supported by the National Science Foundation grant 2325313.\\
The authors acknowledge extensive contributions by Ronald Caplan to the EPREM source code, especially to the algorithms used to solve the focused transport equation. This work benefited from previous conversations with Christina Cohen, Gabriel Muro, and Erika Palmerio about spacecraft observations of specific widespread SEP events.
\end{acknowledgements}

\begin{contribution}
MY developed the analysis code, maintains the EPREM-related code repositories, performed the simulation runs, created the figures, and wrote the original manuscript. BP provided the funding and edited the manuscript. Both authors were responsible for developing the initial research concept.
\end{contribution}

\clearpage

\bibliographystyle{aasjournal}
\bibliography{main.bib}

\appendix

\section{Focused Transport Equation (FTE)}
\label{sec:Focused Transport Equation}

The analytic form of the FTE that EPREM solves, which is based on the approach described in \citep{kota_simulation_2005}, follows from the theory developed by \citet{skilling_cosmic_1971} and \citet{ruffolo_effect_1995}. It can be written as

\begin{eqnarray}
    \frac{d f_s}{d t} & \qquad \textrm{(convection)} \nonumber \\
    + v\mu \hat{b} \cdot \nabla f_s & \qquad \textrm{(streaming)} \nonumber \\
    + \frac{\left(1 - \mu^2\right)}{2} \left[- v \hat{b} \cdot \nabla \ln B - \frac{2}{v}\hat{b} \cdot \frac{d \vec{V}}{d t} + \mu\frac{d \ln \left(n^2 / B^3\right)}{d t}\right]\frac{\partial f_s}{\partial \mu} & \qquad \textrm{(adiabatic focusing)} \nonumber \\
    + \left[-\frac{\mu}{v}\hat{b} \cdot \frac{d \vec{V}}{d t} + \mu^2\frac{d \ln \left(n/B\right)}{d t} + \frac{\left(1 - \mu^2\right)}{2}\frac{d \ln B}{d t}\right]\frac{\partial f_s}{\partial \ln p} & \qquad \textrm{(cooling)} \nonumber \\
    = \frac{\partial}{\partial \mu}\left(\frac{D_{\mu\mu}}{2}\frac{\partial f_s}{\partial \mu}\right) \,, & \qquad \textrm{(pitch-angle scattering)}
    \label{eqn:focused-transport}
\end{eqnarray}

\noindent where $f_{s}\left(t, \vec{r}, p, \mu\right)$ is the distribution of species $s$, $t$ is time, $p$ is momentum, $\mu$ is the pitch-angle cosine, $\vec{B}$ is the magnetic field, $\vec{V}$ is the large-scale velocity field, $n$ is the plasma density, $\hat{b}$ is a unit vector parallel to $\vec{B}$, $v$ is the particle velocity (distinct from $\vec{V}$), and $D_{\mu\mu}$ is the pitch-angle diffusion tensor. 

EPREM computes the distribution function, $f_{s}$, at the location of each node at each time step. By default, EPREM outputs $f_{s}$, which the user may convert to particle flux, $J_{s}\left(t, E\right)$, via the relation
\begin{equation}
    J_{s}\left(t, E\right) = \frac{2Ef(t, E)}{m_{s}^{2}}
\end{equation}
It is also possible to force EPREM to directly compute $J_{s}$ from $f_{s}$ by setting \texttt{outputFlux=1} at runtime.

In addition to modeling field-aligned ion transport and acceleration by solving Equation \ref{eqn:focused-transport}, EPREM is capable of modeling perpendicular drift and diffusion at the user's request. Both features involve additionally solving the a convection--diffusion equation, as described in detail by \citet{kozarev_modeling_2010}. That equation takes the form
\begin{equation}
    \frac{\partial f_{s0}}{\partial t} = \nabla\cdot\left(\kappa_{\perp}\nabla f_{s0}\right) - \vec{v}_{D}\cdot\nabla f_{s0}
    \label{eqn:convection-diffusion}
\end{equation}
where $f_{s0}$ is the average distribution function for species $s$, $\kappa_{\perp}$ is the coefficient of perpendicular diffusion, and $\vec{v}_{D}$ is the particle drift velocity. The latter is given by
\begin{equation}
    \vec{v}_{D} = \frac{cvp}{3q}\nabla \times \left(\frac{\vec{B}}{|\vec{B}|^2}\right)
    \label{eqn:drift-velocity}
\end{equation}
where $v$ is the particle speed, $p$ is the particle momentum, $q$ is the particle charge, and $\vec{B}$ is the local magnetic field.

\section{Lagrangian Grid}
\label{sec:Lagrangian Grid}

EPREM manages a set of surfaces on which it places nodes for calculation. There are six computational surfaces that are logically arranged as the faces of a cube. However, the edges of these faces define logical connections only --- not rigid boundaries (e.g., as on a Euclidian cube). The outward movement of computational nodes effectively deforms the cube of logical faces into a cubed sphere \citep[e.g., Figure 1 of][]{sadourny_conservative_1972}. Each node advances according to $\delta \vec{r} = \vec{V}\delta t$, where $\vec{r}$ is the 3D node displacement, $\vec{V}$ is the 3D flow velocity, and $\delta t$ is the time step duration. It then calculates $f_{s}$ according to the FTE. Each set of linked nodes (typically called a ``stream'') defines a velocity path line that tracks the trajectory of the distribution of particles. In purely steady-state conditions, these velocity path lines represent formal streamlines. If, and only if, the frozen-in condition of MHD holds, they also represent magnetic field lines. EPREM creates the initial grid (known as ``seeding'' the nodes) by generating a new node on every stream, pushing the node to $r = V_{r}\delta t$, and rotating the inner boundary by $\Delta \phi = \Omega_{\odot} \delta t$, where $V_{r}$ is the radial component of the solar wind and $\Omega_{\odot}$ is the solar rotation rate. It repeats this process until it has seeded all nodes. The default configuration of EPREM streams is therefore a Parker spiral with the outermost node at $R_{\text{max}} = R_{\text{min}} + N V_{r}\delta t$, where $R_{\text{min}}$ is the location of the inner boundary (typically the solar surface) and $N$ is the number of nodes. By nature of the Lagrangian grid, EPREM does not have a fixed outer boundary.

\begin{figure}
    \centering
    \includegraphics[width=1.0\textwidth]{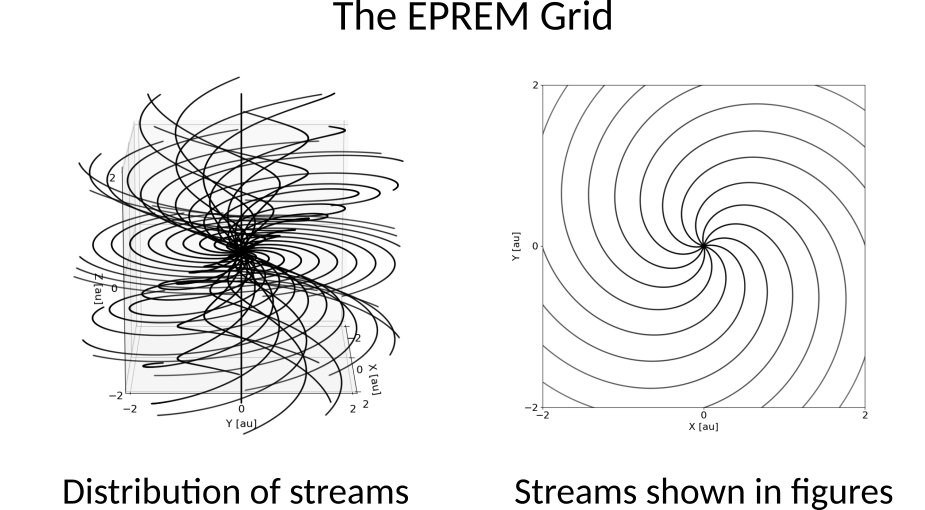}
    \caption{The EPREM grid of nodes. \textit{Left:} a subset of node streams representing the full grid. \textit{Right:} the subset of streams shown in figures such as Figure \ref{fig:stream-flux-baseline}.}
    \label{fig:placeholder}
\end{figure}

In general, EPREM requires knowledge of $\vec{B}$, $\vec{V}$, and $n$ in order to solve Equation \ref{eqn:focused-transport}; for this reason, it has historically been one-way coupled to MHD models such as BATS-R-US, ENLIL, and MAS. The simulation runs presented in this work come from ``uncoupled'' EPREM, which internally generates the requisite MHD values based on configurable parameters for a given simulation run, as described next.

\section{Ideal Shock}
\label{sec:Ideal Shock}

EPREM is capable of generating values of $\vec{B}$, $\vec{V}$, and $n$ for each node, at each time step, from analytic forms. The most basic analytic specification of the magnetic field is that of a Parker spiral. This, combined with a constant radial velocity and a density that decreases as $n(r) \propto r^{-2}$, comprise the default MHD environment regardless of coupling to external models.

The EPREM ideal shock model modifies the default magnetic-field components, velocity-field components, and density to simulate a simple radially propagating shock. The density jump of an EPREM ideal shock follows the standard Rankine-Hugoniot relation $n_{1} = q_{s} n_{2}$, where subscripts 1 and 2 denote downstream and upstream quantities, respectively, and $q_{s}$ is the compression ratio. The form of the background (i.e., unshocked) density is $n(r) = n_{0}\left(\text{1 au}/r\right)^{2}$, where $n_{0}$ is the density at 1~au. The user may set $q_{s}$ and $n_{0}$ at runtime.

In the shock frame of reference, the radial speed of the solar wind \emph{decreases} from $u_{r1}$ to $u_{r2}=u_{r1}/q_{s}$ across the ideal shock front while the remaining two velocity components, which are identically zero in this simplified model, do not change. Transforming to the inertial simulation frame is simply a matter of subtracting the radial speed of the shock, $V_{s}$, from $u_{r1}$ and $u_{r2}$. This leads to a shocked radial wind speed of
\begin{eqnarray}
    V_{r2} &=& V_{s} + \frac{V_{r1} - V_{s}}{q_{s}} \nonumber \\
           &=& \frac{V_{r1}}{q_{s}} + V_{s}\left(1 - \frac{1}{q_{s}}\right) \nonumber
\end{eqnarray}
Both $V_{s}$ and the unshocked radial wind speed, $V_{r1} = u_{r1} - V_{s}$, are input parameters.

The solar-wind magnetic field components are given by
\begin{eqnarray}
    B_{r} &=& B_{0}\left(\frac{\text{1 au}}{r}\right)^{2} \nonumber \\
    B_{\theta} &=& 0.0 \nonumber \\
    B_{\phi} &=& -r B_{r}\left(\frac{\Omega_{\odot}\sin\theta}{V_{r}}\right) \nonumber
\end{eqnarray}
where $\Omega_{\odot}$ is the angular frequency of the Sun's rotation.

Since the shock-normal direction is exactly the radial direction (by definition of our simple model), $B_{r}$ does not change across the shock front, and the coplanarity theorem dictates that $B_{\theta}$ must also not change across the shock front. Finally, applying the Rankine-Hugoniot relations to $B_{\phi}$ leads to $B_{\phi2} = B_{\phi1}\left(u_{r1}/u_{r2}\right) = B_{\phi1} q_{s}$.

By default, the ideal shock produces a step function in the MHD quantities. However, it is possible to make the variation in $n$, $V_{r}$, and $B_{\phi}$ at the shock front more gradual and to cause all shocked quantities to asymptotically return to their upstream values at an exponential rate of the form $Q \propto \exp\left[\xi_{s}\left(r - V_{s}\Delta t\right)\right]$.

\section{Observers}
\label{sec:Observers}

There are two types of ``observers'' in an EPREM simulation run: stream observers and point observers. Each EPREM node records, among other data, the magnetic field, velocity field, density, and ion distribution at each time step. Since streams of linked nodes constitute the EPREM grid, these stream observers provide the fundamental simulated observations. In conjunction with each node's spatial coordinates, it is possible to dynamically track the evolution of any simulated quantity (or any physical quantity derived from the simulated quantities) throughout the simulation volume. The obvious trade-off is that knowledge of any physical quantity as a function of position within the simulation volume is subject to the location of nodes at a given time step.

EPREM point observers attempt to provide a record of simulated quantities at fixed locations within the simulation volume by interpolating values from the nearest nodes at each time step. Each point observer interpolates physical quantities from nearby nodes according to a function of the form $| r_{j}^{\prime} |^{p}$, where $r^{\prime}$ is the distance to the $j^{th}$ node and the value of $p$ is set by the runtime parameter \texttt{interpWeight}. By default, EPREM will interpolate each simulated quantity from all nodes; the user may set the maximum allowed value of $r^{\prime}$ by passing a value (in au) to \texttt{interpDistance}. The contrast between the Lagrangian simulation grid and fixed (i.e., Eulerian) point observers means that some care must be taken in choosing values of \texttt{interpWeight} and \texttt{interpDistance}, as well as when interpreting the time series of any physical quantity recorded by each point observer.

\section{Background Spectrum}
\label{sec:Background Spectrum}

EPREM initializes the distribution of ion species $s$ from a flux spectrum of the form given by Equation \ref{eqn:initial-spectrum}, which it converts to an isotropic particle distribution.

\begin{equation}
    J_{s}(E, r) = \alpha_{s}J_{0}\left(\frac{r}{r_{0}}\right)^{-\beta}\left(\frac{E}{E_{0}}\right)^{-\gamma}\exp\left(-\frac{E}{E_{c}}\right)
    \label{eqn:initial-spectrum}
\end{equation}
where $E$ is the ion energy, $r$ is the radial distance from the inner boundary, $\alpha_{s}$ is the abundance of species $s$ relative to that of protons, $J_{0}$ is the flux at $r_{0}$ of protons with energy $E_{0}$, $E_{c}$ is the roll-over (or cutoff) energy, and $\beta, \gamma \in \mathbb{R}$. Although it is possible to simulate multiple ion species in the same run, EPREM currently does not include any interaction among ion species.

\end{document}